%% file: main.tex
\documentclass[conference,compsoc]{IEEEtran}
\pdfoutput=1

\newcommand{\papertitle}[0]{Catching Worms, Trojan Horses and PUPs: Unsupervised Detection of Silent Delivery Campaigns}

\newcommand{\system}[0]{Beewolf\xspace}
\newcommand{\graph}[0]{galaxy\xspace}

\usepackage{combelow}
\usepackage{amsmath}
\usepackage{amssymb}
\usepackage[normalem]{ulem}  	%
\usepackage{soul}				%
\usepackage{xcolor}				%
\usepackage{url}             			%
\usepackage{cite}            			%
\usepackage[%
  font=footnotesize, 
  caption=true]{subfig}     			%
\usepackage{calc}            			%
\usepackage{shadow}          		%
\usepackage{fancyvrb}        		%
\usepackage{xspace}          		%
\usepackage{ifthen}          			%
\usepackage{comment}         		%
\usepackage{multirow}        		%
\usepackage{dcolumn}         		%
\usepackage{colortbl}
\usepackage{tabularx}        		%
\usepackage{mdwtab}          		%
\usepackage{mdwlist}			%
\usepackage{ifpdf}           			%
\usepackage{rotating}        		%
\usepackage{listings}
\usepackage{wrapfig}         		%
\usepackage{pdfpages}			%
\usepackage{graphicx}        		%
\usepackage{mfirstuc}			%
\usepackage{fixltx2e}
\usepackage[%
  breaklinks=true,
  pdfkeywords={security, vulnerabilities, cyber attacks, software updates, 
  	telemetry, WINE, empirical studies}]{hyperref}
\usepackage{float}
\usepackage{algorithm}
\usepackage{algpseudocode}
\makeatletter
\def\BState{\State\hskip-\ALG@thistlm}
\makeatother

\makeatletter
\def\@copyrightspace{\relax}
\makeatother

\usepackage[table, subgroups]{svn-multi} 

\svnidlong
{$HeadURL: https://mc2.umiacs.umd.edu/svn/SDSatUMD/papers/2016_lockstep_behavior/lockstep_behavior.tex $}
{$LastChangedDate: 2016-08-15 13:08:53 -0400 (Mon, 15 Aug 2016) $}
{$LastChangedRevision: 3918 $}
{$LastChangedBy: tdumitra $}

\def\paperversionmajor{1}          %
\def\paperversionminor{\svnrev}    %

\def\monthName#1{\ifcase#1\or
  January\or February\or March\or April\or May\or June\or
  July\or August\or September\or October\or November\or December\fi}

\hypersetup{pdftitle={\papertitle}}
\hypersetup{pdfproducer={v. \paperversionmajor.\paperversionminor}}
\ifpdf
\fi
\renewcommand{\paragraph}{\vspace{3pt}\noindent\textbf}
\sethlcolor{lime}

\newcounter{hypothesis}                                     %

\newcounter{finding}                                     %

\newcommand{\topic}[1]{\smallskip \noindent{\bf #1.}}
\newcommand{\eat}[1]{} %

\newcommand{\squishlisttwo}{
 \begin{list}{$\bullet$}
  { \setlength{\itemsep}{2pt}
     \setlength{\parsep}{0pt}
    \setlength{\topsep}{0pt}
    \setlength{\partopsep}{0pt}
    \setlength{\leftmargin}{1.0em}
    \setlength{\labelwidth}{1em}
    \setlength{\labelsep}{0.5em} } }

\newcommand{\squishend}{
  \end{list}  }

\begin{document}

\date{}

\title{\papertitle}

\author{
\IEEEauthorblockN{Bum Jun Kwon}
\IEEEauthorblockA{University of Maryland\\
bkwon@umd.edu}
\and
\IEEEauthorblockN{Virinchi Srinivas}
\IEEEauthorblockA{University of Maryland\\
virinchi@cs.umd.edu}
\and
\IEEEauthorblockN{Amol Deshpande}
\IEEEauthorblockA{University of Maryland\\
amol@cs.umd.edu}
\and
\IEEEauthorblockN{Tudor Dumitra\cb{s}}
\IEEEauthorblockA{University of Maryland\\
tdumitra@umiacs.umd.edu}
}

\maketitle

\pagenumbering{arabic}

\input{numbers_arxiv}
\input{abstract_arxiv}
\input{intro_arxiv}
\input{problem_arxiv}
\input{data_arxiv}
\input{locksteps_arxiv}
\input{campaigns_arxiv}
\input{related_arxiv}
\input{conclusions_arxiv}

{
\bibliographystyle{abbrv}

\small
\bibliography{security_arxiv}
\input{pseudocode}
}

\end{document}

%% file: numbers_arxiv.tex
\newcommand{\NumLockSteps}[0]{127,495\xspace}
\newcommand{\NumLockStepsApprox}[0]{130,000\xspace}
\newcommand{\NumLockStepsVerDlr}[0]{67,094\xspace}
\newcommand{\NumLockStepsVerUrl}[0]{60,401\xspace}
\newcommand{\NumCampaignsMillions}[0]{1.4\xspace}
\newcommand{\NumEvents}[0]{33,265,048\xspace}
\newcommand{\NumEventsMillions}[0]{33.3\xspace}
\newcommand{\NumPayloads}[0]{673,903\xspace}
\newcommand{\NumPayloadsMillions}[0]{0.7\xspace}
\newcommand{\NumHosts}[0]{1,869,456\xspace}
\newcommand{\NumHostsMillions}[0]{1.9\xspace}
\newcommand{\NumDownloaders}[0]{83,088\xspace}
\newcommand{\NumDomains}[0]{60,002\xspace}
\newcommand{\HashInVTRatio}[0]{17\%\xspace}
\newcommand{\AdwareRateThresholdAdware}[0]{10\%\xspace}
\newcommand{\DetectionRateThresholdMalware}[0]{30\%\xspace}
\newcommand{\NumMalware}[0]{1,228\xspace}
\newcommand{\NumAdware}[0]{15,350\xspace}
\newcommand{\NumDataPoints}[0]{121\xspace}
\newcommand{\CutLevel}[0]{7\xspace}

%% file: abstract_arxiv.tex
\begin{abstract}
\noindent 
The growing commoditization of the underground economy has given rise to malware delivery networks,
which charge fees
for quickly delivering 
malware or unwanted software 
to a large number of hosts.
To provide this service, a key method is 
the orchestration of silent delivery campaigns, which involve a group of downloaders that receive remote commands and that deliver their payloads without any user interaction.
These campaigns have not been characterized systematically, unlike other aspects of malware delivery networks. 
Moreover, silent delivery campaigns can evade detection by relying on inconspicuous downloaders on the client side 
and on disposable domain names on the server side.
We describe \system, a system for detecting silent delivery campaigns from Internet-wide records of download events. 
The key observation behind our system is that the downloaders involved in these campaigns frequently retrieve payloads in lockstep.
\system identifies such locksteps in an unsupervised and deterministic manner. 
By exploiting novel techniques and empirical observations, \system can operate on streaming data.
We utilize \system to study silent delivery campaigns at scale, on a data set of \NumEventsMillions million download events.
This investigation yields novel findings, e.g.
malware distributed through compromised software update channels, 
a substantial overlap between the delivery ecosystems for malware and unwanted software, and
several types of business relationships within these ecosystems.
\system 
achieves over 92\% true positives and fewer than 5\% false positives.
Moreover, \system can detect suspicious downloaders a median of 165 days ahead of existing anti-virus products 
and payload-hosting domains a median of 196 days ahead of existing blacklists.

\end{abstract}

%% file: intro_arxiv.tex
\vspace{-3pt}
\section{Introduction}
\label{sec:intro}
\vspace{-3pt}
The growing commoditization of the underground economy has given rise to malware delivery networks~\cite{DBLP:conf/sp/LiAXYW13, caballero2011measuring}, 
which orchestrate \emph{campaigns} to quickly deliver 
malware 
to a large number of hosts.
Understanding these campaigns can provide new insights into the malware landscape. 
For example, the ability to measure the duration of such campaigns would reveal 
which malware families remain active and which are likely to stop propagating. 
By tracking the \emph{downloaders} and the \emph{domain names} associated with each malware delivery campaign, and the malware \emph{payloads} disseminated, we could infer 
the business relationships from the underground economy.
Establishing precise time bounds for the campaigns would also enable correlation with other concurrent events, such as additional activities and downloads performed by the malware samples delivered within each campaign. 
This new understanding has the potential to expose fragile dependencies in the underground economy, leading to effective intervention strategies for disrupting the malware delivery process~\cite{thomas2015framing}.

Prior work has generally focused on identifying the malicious domains 
\cite{DBLP:conf/uss/AntonakakisPDLF10, %
	DBLP:conf/ndss/BilgeKKB11,%
	DBLP:conf/uss/AntonakakisPLVD11,%
	DBLP:conf/sp/LiAXYW13,%
	DBLP:conf/esorics/ManadhataYRH14,%
	DBLP:conf/icdcs/ZhangSGLM15,%
	DBLP:conf/uss/NelmsPAA15,%
	DBLP:conf/dsn/RahbariniaPA15}, 
the malware families disseminated 
\cite{DBLP:conf/raid/CovaLTKD10,%
	Vadrevu13:MalwareDownloads,%
	Invernizzi14:Nazca,%
	Kwon15:DropperEffect,%
	kotzias2015pup} 
and, to a lesser extent, the downloaders utilized on the client side~\cite{	Kwon15:DropperEffect}.
Comparatively less attention has been given to the task of precisely characterizing 
the relationships among these entities;
for example, a comprehensive ground truth about past malware delivery campaigns is currently unavailable. 
As a step toward understanding campaigns, we focus on a particular subset called \emph{silent delivery campaigns}, which involve a group of downloaders that receive remote commands and that download their payloads with no user interaction. 
These campaigns are particularly attractive to the organizations that disseminate malware 
or potentially unwanted programs (PUPs),
as they can evade detection by utilizing inconspicuous downloaders, to retrieve the payloads, and disposable domain names, to host and serve it temporarily. 
We propose unsupervised and deterministic techniques for detecting silent delivery campaigns.
We also describe the design of a system, called \system,%
\footnote{Beewolves are a species of wasp that hunts bees, which are known to exhibit group behaviors.}
 which implements these techniques and can operate either on the entire data set of download events (\emph{offline mode}) or on a stream of data (\emph{streaming mode}).
Using \system, we conduct the first systematic study of silent delivery campaigns.

The key observation behind \system is that, when 
downloaders across the Internet are instructed to conduct a campaign, they will all access a common set of DNS domains, within a short time window, to retrieve the payloads. 
After a period of inactivity, the same downloaders will request additional payloads from a set of fresh domains.
This \emph{lockstep behavior} exposes the fact that the downloaders are controlled remotely and 
reveals the domains involved in subsequent campaigns.
We expect that we can parametrize lockstep detection to distinguish benign software updates that are initiated remotely and malicious campaigns. 
In particular, software updaters repeatedly access the same server-side infrastructure, while malicious campaigns exhibit a high domain churn as they try to evade blacklists.
Additionally, we can whitelist the known benign updaters to further reduce the false positive rate. 
Our approach is complementary 
to the machine learning techniques proposed for detecting malicious domains \cite{DBLP:conf/raid/CovaLTKD10,Vadrevu13:MalwareDownloads,Invernizzi14:Nazca,DBLP:conf/icdcs/ZhangSGLM15}.
In contrast to these techniques, recognizing a lockstep pattern in a stream of Internet-wide download events yields an intuitive explanation of the underlying activity, without interpreting clusters of events defined by multiple features.

We formulate lockstep detection as a graph mining problem.
We construct a \emph{bipartite graph}, where a node corresponds to either a downloader or a payload hosting domain, and an edge indicates that a downloader contacted a domain to retrieve a payload. 
A lockstep is a \emph{near biclique}%
\footnote{We allow a few edges to be missing to account for download events that are occasionally not recorded by our data collection infrastructure.}
in this graph---a graph component that is almost fully connected, except for a few missing edges---with the added constraint that 
the edges are created 
within a short time window $\Delta t$.
Existing algorithms for lockstep detection \cite{catchsync, copycatch, SynchroTrap} are not well suited for finding 
silent delivery campaigns
because they require seed nodes 
to bootstrap the algorithm and because they 
are not designed to operate on streaming data.
In contrast, downloaders typically remain undetected for several months~\cite{Kwon15:DropperEffect}, making it difficult to identify seeds in a timely manner, and malicious domains can be 
discarded within days~\cite{DBLP:conf/raid/KuhrerRH14, Thomas16:CommercialPPI}, 
at which point the information from lockstep detection is no longer actionable.

We propose a
novel lockstep detection technique, which can operate on streams of download events.
We perform the computationally intensive operations (e.g., updating the bipartite graph and the adjacency lists) incrementally, as new events are received, and then we detect locksteps using an efficient linear algorithm. 
We use this technique in both of \system's modes of operation. 
In \emph{offline mode}, \system analyzes our entire download events, with the aim of characterizing lockstep behaviors empirically.
In \emph{streaming mode}, \system receives data incrementally and prunes the locksteps detected to 
focus on suspicious downloaders and domains.

We utilize \system to conduct a large empirical study of silent delivery campaigns conducted over one year.
We analyze a data set of \NumEventsMillions million download events, observed on \NumHostsMillions million hosts, and we detect over \NumLockStepsApprox locksteps. These locksteps comprise \NumCampaignsMillions million campaigns.
Building on the observation that many downloaders involved in lockstep behavior have valid digital signatures, we identify \emph{representative publishers} for each lockstep and we analyze the relationships among publishers.
This investigation yields new insights into silent delivery campaigns, including 
malware distributed through compromised software update channels, 
a substantial overlap between the malware and PUP delivery ecosystems, and
several types of business relationships within these ecosystems.
We also show that \system 
achieves over 92\% true positives and fewer than 5\% false positives, 
and that it can detect suspicious downloaders a median of 165 days ahead of existing anti-virus products 
and payload-hosting domains a median of 196 days ahead of existing blacklists. 
Building on our empirical investigation, we implement several optimizations that allow \system to operate on streaming data.

In summary, we make the following contributions: 
\begin{enumerate}
\item We conduct a systematic study of malware delivery campaigns and we report several new findings about the malware and PUP delivery ecosystems.
\item We propose techniques for discovering silent delivery campaigns by detecting lockstep behavior in large scale collections of download events. These techniques are unsupervised and deterministic, as they do not require seed nodes and are not based on machine learning. 
\item We present a system, \system, which implements these techniques, along with evidence-based optimizations that allow it to detect silent delivery campaigns in a streaming fashion. 

\end{enumerate}

This paper is organized as follows: In section \ref{sec:problem}, we characterize the threat of silent delivery campaigns and we state our goals. 
We describe our data set and the methods we use for distinguishing between malware and PUPs in Section \ref{sec:datasets}. We discuss the key components of \system in Section \ref{sec:lockstepdetection}. 
In Section \ref{sec:campaigns}, we characterize silent delivery campaigns. In the following sections, we evaluate the performance of \system. Section \ref{sec:detection-performance} presents the detection performance and in Section \ref{sec:evaluation} we evaluate the performance of \system's streaming mode. 
We review related work in section \ref{sec:related-work}.

%% file: problem_arxiv.tex
\vspace{-3pt}
\section{Threat model}
\label{sec:problem}
\vspace{-3pt}

Downloader trojans (also known as droppers) are at the heart of malware distribution techniques~\cite{Kwon15:DropperEffect}. 
A \emph{downloader} is an executable program that connects to an Internet \emph{domain} and downloads other executables (called \emph{payloads}), usually in response to remote commands. 
We focus on the domains hosting the payloads, which are often distinct from other components of the malware delivery networks, e.g. exploit servers, command \& control servers, payment servers~\cite{DBLP:conf/ccs/XuNBYCG14}, and we take only the second level domain (SLD) under a public suffix\footnote{We use Mozilla's public suffix list from \url{https://publicsuffix.org/}.} (e.g., \verb!site1.com!, \verb!site2.co.uk!).

\topic{Silent delivery campaigns}
Malware delivery networks use a variety of methods to install their downloaders, e.g. drive-by-download exploits, social engineering, affiliate programs~\cite{caballero2011measuring}.
When they receive new payloads from their clients, the malware delivery networks command their downloaders to retrieve these payloads on the victim hosts. 
This results in coordinated waves of payload delivery, which often do not require any user intervention to avoid attracting attention.
We term these waves \emph{silent delivery campaigns}, by analogy with the silent updating mechanisms increasingly adopted by benign software publishers~\cite{Dubendorfer09:BrowserSecurity, Nappa15:VulnerabilityPatching}.
A key difference between the silent delivery campaigns conducted on behalf of malicious and benign payloads is that benign campaigns repeatedly access the same server-side infrastructure, while malicious campaigns exhibit a high domain churn as they try to evade blacklists. 
Depending on the payloads, these campaigns may be \emph{malware delivery campaigns}, which drop executables with unambiguously malicious functionality such as trojan horses, bots, keystroke loggers, or \emph{PUP delivery campaigns}, which drop PUPs such as adware, spyware and even additional droppers.

The detection of domains involved in malware and PUP distribution has been widely explored using machine learning techniques~\cite{DBLP:conf/raid/CovaLTKD10,Vadrevu13:MalwareDownloads,Invernizzi14:Nazca,DBLP:conf/icdcs/ZhangSGLM15}. 
These techniques typically output clusters of events, defined by multiple features, which can be difficult to interpret. 
We investigate a complementary approach: deterministic techniques, based on the intuition that \emph{temporal patterns in the downloader-domain interactions can expose remotely controlled downloaders}.

\topic{Lockstep behavior}
The coordinated downloads from silent delivery campaigns result in \emph{lockstep behavior}.
Intuitively, lockstep behavior corresponds to repeated observations of synchronized activity among 
a group of downloaders (or domains), which 
access (are accessed by)
the same set of 
domains (downloaders) to retrieve payloads,
within a bounded time period. 
In other words, locksteps capture coordinated downloads that do not experience random delays, e.g. from manual user intervention. 
This points to silent delivery campaigns. 
As lockstep detection requires several repeated observations of coordinated downloads, \emph{a lockstep may correspond to one or several delivery campaigns that use the same infrastructure.}

Formally, 
consider a bipartite graph $G=(U,V,E)$ where $U$ and $V$ are disjoint set of nodes corresponding to \emph{left hand nodes} and \emph{right hand nodes}, respectively, and an edge $e \in E$ may link two nodes belonging to different sets but not nodes from the same set. Let $t_{i,j}$ represent the time at which an edge is formed between node $i \in U$ and node $j \in V$. Further, let $U' \subseteq U$ and $V' \subseteq V$. We define a \textit{star} $[U',{j},\Delta t, \delta t]$ on $U'$ and some central node $j \in V'$ as follows:
\begin{align}
   \mid U' \mid \geq 2
   \label{eq:min_star_condition} 
\end{align}
\vspace{0.1pt}
\begin{align}
(\max\limits_{i} t_{i,j} -  \min\limits_{i} t_{i,j}) \leq \Delta t \;  \forall i \in U'
	\label{eq:star_condition} 
\end{align}
The above equations state that a star contains at least $2$ left hand nodes and the time difference between the addition of the first and the last edge to the star is at most $\Delta t$.

A \textit{lockstep} $[U',V',\Delta t,\delta t]$ in $G(U,V,E)$ satisfies the following constraints:
\begin{align}
   \mid U' \mid > 2
\end{align}
\vspace{0.1pt}
\begin{align}
\exists V_i^{'} \subseteq V'   \;\; \forall i\;\; \mbox{s.t.} \;\; \mid V_i^{'}  \mid > 2\;\; \mbox{and}\;\;\mid V_i^{'} \mid \geq \alpha \mid V' \mid 
\end{align}
\vspace{0.1pt}
\begin{align}
(i,j) \in E \;\; \forall i \in U',\;j \in V_i^{'}
\end{align}
The above equations specify that a lockstep contains more than $2$ nodes each from $U$ and $V$ and that the subgraph induced by these nodes is nearly complete.
If $\alpha=1.0$, this subgraph is a \textit{complete biclique}, while for any value  $\alpha_{min} \leq \alpha < 1$, the lockstep corresponds to a \textit{near-biclique}. 
Such a near or complete biclique represents a \emph{lockstep} if it also satisfies the following temporal constraints,
for a predefined $\Delta t$ and $\delta t$ and for 2 distinct stars defined on $j,j' \in V_{i'}$: %
\begin{align}
(\max\limits_{i} t_{i,j} -  \min\limits_{i} t_{i,j}) \leq \Delta t \;  \forall i \in U'
\end{align}
\vspace{0.1pt}
\begin{align}
(\max\limits_{i} t_{i,j'} -  \min\limits_{i} t_{i,j'}) \leq \Delta t \;  \forall i \in U'
\end{align}
\vspace{0.1pt}
\begin{align}
\mid \max\limits_{i} t_{i,j'} - \max\limits_{i} t_{i,j} \mid \geq \delta t
\end{align}

The above temporal constraints ensure that a lockstep contains at least 2 stars that are at least $\delta t$ apart in time. Further, if the same star occurs in multiple timestamps, we consider it only once inside a lockstep.
\begin{figure*}[!htb]
\vspace{-10pt}
\centering
\includegraphics[width=150mm]{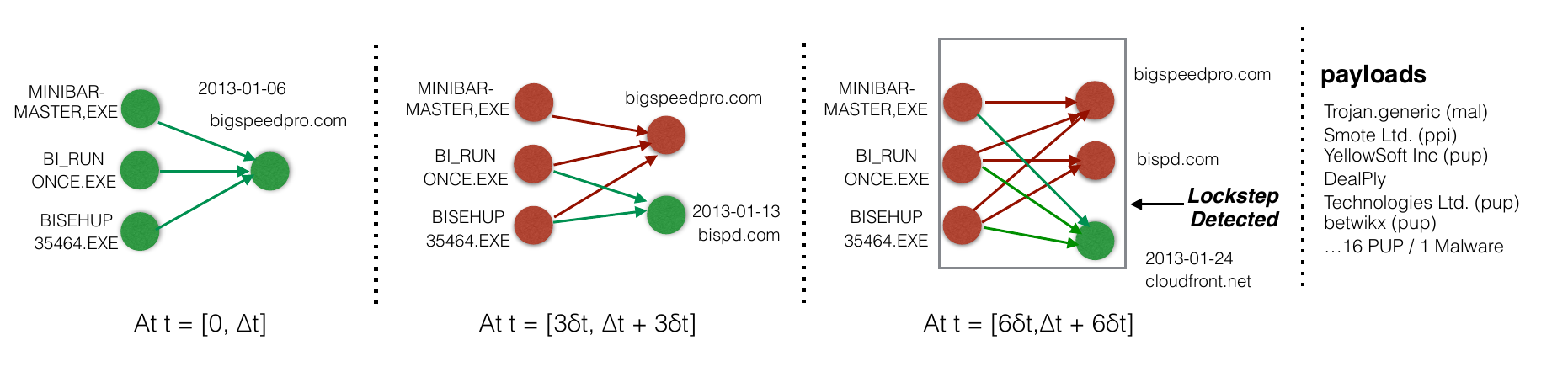}
\vspace{-10pt}
\caption{Lockstep Illustration (Red color corresponds to existing nodes and edges. Green color corresponds to new nodes and edges which we receive in the data stream in an online fashion).}%
\vspace{-10pt}
\label{fig:lockstep_fig}
\end{figure*}
We illustrate the lockstep behavior in Figure~\ref{fig:lockstep_fig}. 
For the problem of detecting silent delivery campaigns, the nodes of the bipartite graph correspond to downloaders and domains.
There is an edge between a domain and a downloader in the bipartite graph if the downloader accessed the domain to drop a payload. 
The payload information is captured as an attribute on the edge.
A star can have (i) multiple downloaders accessing the same domain; or (ii) multiple domains being accessed by a single downloader. The formal definition of a star (equation \ref{eq:star_condition}) considers a $j$ which could be either a downloader or a domain. Note that a single edge does not count as a star because of condition~(\ref{eq:min_star_condition}).
Having these two different star topologies help us detect behaviors such as (i) campaign changing to a different domain after a C\&C server takedown, (ii) domains within a same campaign establishing connection with a new version of downloader. 
In Figure~\ref{fig:lockstep_fig}, at time $t=0$, we observe a star with 3 downloaders accessing a domain. 
At $t=3\delta t$, although we observe new stars, they do not correspond to a lockstep as a lockstep must contain more than $2$ domains and $2$ downloaders according to our lockstep definition. 
Then, at $t=6\delta t$, we observe a near-biclique,
with $\alpha \ge 0.8$, that we detect as a lockstep. 
We can observe that a lockstep corresponds to a series of campaigns. The lockstep consists of set of stars across different time windows. We exploit the gap between these time windows, and define a \emph{campaign} as follows. The activities appearing in the time windows with a gap less than $n\delta t$ will be considered as a single campaign. If the gap is larger than $n\delta t$, we treat it as a different campaign. 

\topic{Streaming}
We adopt the terminology from~\cite{mondal2014stream} and define a stream processing task as a query that is submitted once by the user and is \emph{executed continuously or periodically} by the system, as updates arrive. 
The temporal scope of the task may be either a \emph{sliding window} or the entire \emph{current state of the graph}; 
lockstep detection falls in both these categories, as star detection considers new download events that are \emph{at most} $\Delta t$ apart and a lockstep requires two or more stars that \emph{at least} $\delta t$ apart.
The lockstep detection task is a \emph{quasi-continuous query} that must produce or update a result when the user requests it (once per $\Delta t$), rather than keeping the query result up-to-date whenever the inputs change.

\topic{Adversary model}
A silent delivery campaign will evade detection if its nodes from $U$ do not remain active for at least  $\Delta t + \delta t$ %
or 
if none of these nodes contacts at least 2 nodes from $V$.
For example, payloads make poor choices for nodes in our bipartite graph, as they are frequently repacked and some malware families seek to deliver unique samples to each host~\cite{caballero2011measuring}.
We consider adversaries that have access to some varied, but limited, resources---e.g. downloaders that get updated periodically but not daily (which could raise suspicions), a limited stockpile of domains---so that we can find some values for $\Delta t$ and $\delta t$ that allow us to detect their lockstep behavior.

\topic{Goals}
Lockstep detection is challenging when analyzing large volumes of data.
For example, finding a biclique with the maximum number of edges is an NP complete problem \cite{maximumbiclique}. 
It is also not clear \emph{a priori} how to parametrize lockstep detection in order to distinguish benign software dissemination from malware delivery.

Our \emph{first goal} is to build an efficient and scalable system for detecting lockstep behavior. 
Our system should be unsupervised, i.e., it should not require any prior information or seed nodes. 
The system should be able to operate in real time and to build the locksteps incrementally, as the stream of stars are collected and fed to our system. 
While we evaluate our system using telemetry collected worldwide, similar to data available to security companies, OS vendors, or ISPs, we also aim to lower the deployment bar for small enterprises. 
Specifically, our system should detect locksteps if at least three victims are infected by the same campaign.

Our \emph{second goal} is to conduct a large scale empirical study of silent delivery campaigns. 
These campaigns may deliver benign software, PUPs, malware or a combination of these payload types. 
We aim to illuminate the characteristics and differences among the campaigns conducted by various organizations, and to 
expose the business relationships among these organizations. 
Finally, our \emph{third goal} is to incorporate this domain knowledge into our lockstep detection system and to assess how well it can identify suspicious activity, such as malware or PUP dissemination campaigns.
Using external information about the maliciousness of downloaders and domains caught in locksteps, we aim to assess the true positive and false positive rates\footnote{We cannot estimate the false negative rate because we lack ground truth about malware delivery campaigns. An undetected malicious downloader may be either a false negative or a dropper not controlled remotely.} of this detection system. 
We also aim to measure the lead detection time, compared to the existing sources.

\topic{Non-goals}
We do not aim to detect all possible malware delivery vectors, e.g. download instructions hardcoded into the droppers, malware and PUPs distributed through software bundles, vulnerability exploits, or other mechanisms that do not involve remotely controlling a group of downloaders. 
Our campaigns do not aim to capture the end-to-end attack kill chain and do not include activities performed by the payloads on the hosts where they were downloaded.
Finally, our system should detect silent delivery campaigns in a deterministic manner, without using machine learning.

%% file: data_arxiv.tex
\vspace{-3pt}
\section{Data sets}
\label{sec:datasets}
\vspace{-3pt}

In this section, we describe 
our data sets, our ground truth and our method for distinguishing malware from PUPs. 

\vspace{-3pt}
\subsection{Data Sources}
\vspace{-3pt}
We utilize a large data set of download events, collected by Kwon et al.\cite{Kwon15:DropperEffect}.
These events were reconstructed from observations on end hosts. 
From this data we utilize the SHA2 hash of the downloader and the downloaded file (payload), the source domain of the download, 
and the timestamp of the event.
We focus on events from 2013, as the data set has good coverage for that year.
We exclude the downloads performed by Web browsers, which typically involve user actions. 
We identify the top 5 browsers in our data set by searching the digital signatures for the following $<$publisher, product$>$ pairs: $<$Microsoft Corporation, Internet Explorer$>$, $<$Google Inc, Chrome$>$, $<$Mozilla Corporation, Firefox$>$, $<$Apple Inc, Safari$>$, $<$Opera Software, Opera$>$. Table~\ref{tab:statistics} summarizes our data after this filtering step. 

\begin{table}[t]
\begin{center}
\scalebox{1.0}{
\begin{tabular}{p{4.5cm}p{3cm}}
    \hlx{hv}
Lockstep Behaviors & \NumLockSteps \\
~~$type_{dlr:dom}$        & \NumLockStepsVerDlr    \\
~~$type_{dom:dlr}$       	& \NumLockStepsVerUrl    \\
Total Downloaders     & \NumDownloaders      \\
Domains accessed & \NumDomains \\
Download events  & \NumEventsMillions million\\
Total Payloads	& \NumPayloadsMillions million \\
Hosts       & \NumHostsMillions million \\
    \hlx{vhv}
\end{tabular}}
\end{center}
\vspace{-5pt}
\caption{Summary of our data sets of the year 2013.}
\vspace{-10pt}
\label{tab:statistics}
\end{table}

\vspace{-3pt}
\subsection{Ground Truth Data}\label{sec:groundtruth}
\vspace{-3pt}
While ground truth for malware delivery campaigns is currently unavailable, 
we collect ground truth about executables from multiple sources. 
\topic{VirusTotal} VirusTotal\footnote{\url{https://www.virustotal.com/}} 
provides file scan reports for up to 54 anti-virus (AV) products. We query VirusTotal for the hash of each downloader and payload in our data set to obtain its first-seen timestamp, the number of AV products that flagged it as malicious, the AV detection names assigned to it, the total number of AV products utilized for scanning, and the corresponding file signatures. We were able to retrieve reports for about \HashInVTRatio of the binaries from 2013. 
In line with prior work~\cite{Kwon15:DropperEffect, kotzias2015pup}, we set a threshold 
 $r_{mal}\geq$\DetectionRateThresholdMalware and we flag the files that meet the condition.
\label{malware-pup-distinction}
This process selects both malware and potentially unwanted programs (PUPs). 
To further separate them, we search the AV labels given to these samples for the following keywords: ``adware", ``not-a-virus", ``not malicious", ``potentially", ``unwanted", ``pup", ``pua", ``riskware", ``toolbar", ``grayware", ``unwnt", and ``adload"~\cite{kotzias2015pup}. 
We define $r_{pup}$ to be the percentage of AV labels that include one of these keywords.
We consider that a binary is malware if $r_{mal}\geq$\DetectionRateThresholdMalware and $r_{pup}\leq$\AdwareRateThresholdAdware. It is treated as PUP if $r_{pup}>$\AdwareRateThresholdAdware and $r_{mal}\geq$\DetectionRateThresholdMalware.
We identify \NumMalware malware samples and \NumAdware PUPs through this process.

\topic{National Software Reference Library} NSRL\footnote{\url{http://www.nsrl.nist.gov/}} provides a reference data set (RDS) of benign software. 
We collect the MD5 signatures of the applications and their list of publishers. The version of the RDS we use is at 2.52, which was released in April 2015. We consider benign all the executables where either (1) the hash matches or (2) the publisher matches and has a valid signature.

\topic{Information about publishers} 
To identify publishers engaged in the Pay-Per-Install (PPI) business~\cite{caballero2011measuring}, we utilize two lists of PPI providers from underground forums.%
\footnote{\url{http://ppitalk.com/showthread.php/38-List-of-Pay-Per-Install-Companies}}\footnote{\url{http://www.blackhatworld.com/seo/list-of-pay-per-install-ppi-networks.646987/}} 
For other types of publishers, we query the Reason Labs knowledge base.\footnote{\url{https://www.reasoncoresecurity.com/knowledgebase.aspx}}
This service provides details about the publisher, e.g. whether it is considered safe or if it uses its certificates to sign PUPs.

%% file: locksteps_arxiv.tex
\vspace{-3pt}
\section{Detecting lockstep behaviors in real-time}
\label{sec:lockstepdetection}
\vspace{-3pt}

In this section, we describe the design and implementation of \system, which detects lockstep behavior in real-time.
\system can operate in two modes. 
In \emph{offline mode}, our system analyzes our entire download events, with the aim of characterizing lockstep behaviors empirically.
We utilize this mode in our experiments from Sections~\ref{sec:campaigns} and~\ref{sec:detection-performance}.
In \emph{streaming mode}, \system receives data incrementally and prunes the locksteps detected to 
focus on suspicious downloaders and domains. 
We evaluate this mode in Section~\ref{sec:evaluation}.
We implement \system in Python, using the \verb!NetworkX!\footnote{\url{https://networkx.github.io/}} package to manipulate graphs.

\begin{figure*}[t]
\vspace{-10pt}
\centering
\includegraphics[width=150mm]{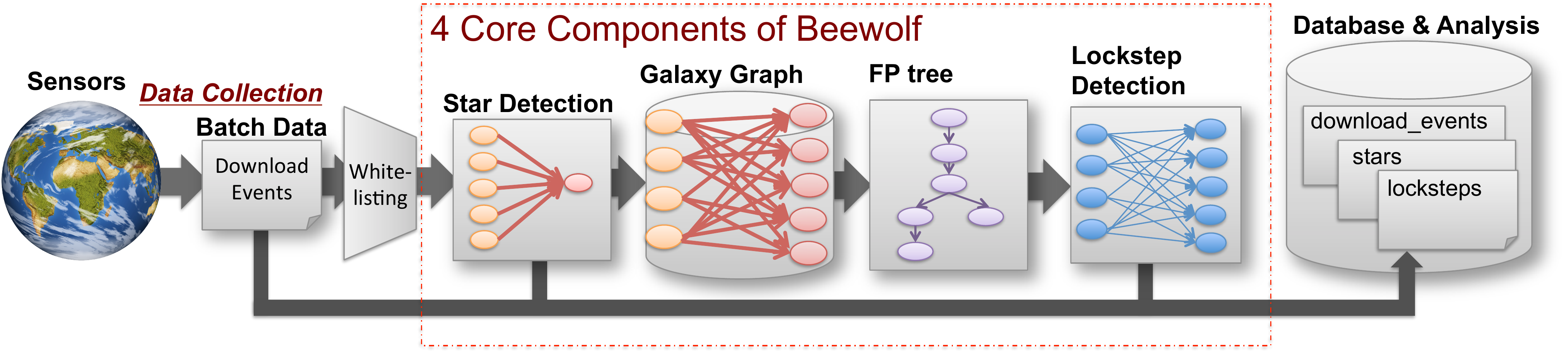}
\vspace{-5pt}
\caption{System architecture.} %
\vspace{-10pt}
\label{fig:system_architecture}
\end{figure*}

As illustrated in Figure~\ref{fig:system_architecture}, \system consists of a data analysis pipeline with four components: star detection, \graph graph construction, frequent pattern (FP) tree construction, and lockstep detection. 
We also maintain a database with three tables: \verb!download_events!, \verb!stars!, and \verb!locksteps!. 
The first step is to detect new \emph{star patterns} 
as new download events are recorded. 
In the rest of the paper, we refer to the bipartite graph as ``\graph graph''. 
The stars detected are updated incrementally in the \graph graph. Further, we traverse the \graph graph to build the FP tree which is an in-memory data structure to detect locksteps.
The algorithm pseudocode of \system can be found at the Appendix.

\vspace{-3pt}
\subsection{Whitelisting}
\label{sec:whitelisting}
\vspace{-3pt}
As discussed in section~\ref{sec:groundtruth}, we identify benign binaries using the NSRL data. We maintain a whitelist, which consists of these benign binaries. Prior to the main data analysis pipeline, we filter out the download events generated by the benign downloaders, which are listed in the whitelist. 
We do not expect this whitelist to be exhaustive---NSRL may not include all the legitimate downloaders---but this simple filtering step helps us focus on the most suspicious campaigns and improves \system's performance.
Moreover, while it is likely unfeasible to whitelist all benign software, only a few programs have a downloader functionality.
Our whitelist contains 6,996 downloaders.

\vspace{-3pt}
\subsection{Star Detection} 
\label{sec:star-detection}
\vspace{-3pt}
Each row of the \verb!download_events! table consists of a downloader ($dlr$), corresponding domain accessed ($dom$), the downloaded file ($payload$), and the timestamp when the download event occurred. 
We assign a unique identifier to each download event in the table, and sort them in ascending time order.
Conceptually, each download event corresponds to an edge in the \graph graph, linking a node represented by $dlr$ with a node represented by $dom$.

Given a moving time window of size $\Delta t$, we 
query the events that occurred within this time range. We utilize these series of download events to identify star patterns. 
We can create stars in two ways, by starting from a downloader and aggregating the adjacent domains, or vice versa.
We assign unique identifiers to each new star, and record the associated events in the \verb!stars! table. 
After generating all the stars within $\Delta t$, we slide the time window by $\delta t$ and repeat the star detection process, until the end of the time window reaches the last event.

\vspace{-3pt}
\subsection{Galaxy Graph} 
\label{sec:bipartite-graph}
\vspace{-3pt}
\system maintains the \graph graph, which has two kinds of nodes: nodes that correspond to downloader programs and nodes that correspond to domains hosting payloads. We represent a node in the \graph graph as $node_{gg}$.
We explicitly maintain only $1$ edge between a downloader and a domain. However, there can arise situations where a downloader accesses a domain at different times; we discuss how we deal with this situation later in this section.

We update the \graph graph incrementally, using the star patterns detected in the previous step. As explained earlier, there are $2$ types of stars. We consider only one type of star and ignore the other while detecting and updating the stars to the \graph graph; \graph graph at any point contains only one type of stars. For simpler explanation, we discuss only the star type corresponding to multiple downloaders accessing the same domain; the same explanation can be extended when dealing with the other star type. Further, we present results corresponding to both star types, when dealt separately in Section~\ref{sec:campaigns}.

When we detect a star, we add the central node (domain) and its adjacent nodes (downloaders accessing it) to the bipartite graph, and we create the corresponding edges. For each newly detected star, while adding the central node (domain) we also specify the star id (e.g. $(2)\; dom_{B}$), in order to separate it from the nodes corresponding to $dom_B$ from different stars. When the new star is a superset of some existing star in the \graph graph, we replace the existing star with it. If it is a subset of some existing star, \system discards it from further processing.

\vspace{-3pt}
\subsection{FP tree} 
\label{sec:fp-tree}
\vspace{-3pt}
We traverse the \graph graph, constructed in the previous step, to build a data structure called a Frequent Pattern (FP) tree. 
The FP tree was used successfully in other domains, for example to design scalable algorithms for frequent pattern mining \cite{han2000mining}.
We employ the FP tree algorithm from \cite{mondal2014eagr}.  
Let us represent a node in the FP tree as $node_{fp}$.
Given the \graph graph $G=(U,V,E)$, 
the algorithm starts by sorting the adjacency list of $V$. 
The adjacency list is a representation of the \graph graph and consists of the collection of neighbor lists for each $node_{gg} \in V$.
The sorting is done in two rounds. 
In the first round, we sort each $node_{gg}$ $v$ $\in$ $V$ by their degree (the number of $v$'s neighbors in $U$), in descending order. 
In the second round we sort each list of neighbors.
Specifically, we sort the neighbors $u$ of $v$ by their degree (the number of $u$'s neighbors in $V$), also in descending order. 

\begin{figure}[t]
\vspace{-5pt}
\centering
\includegraphics[width=85mm]{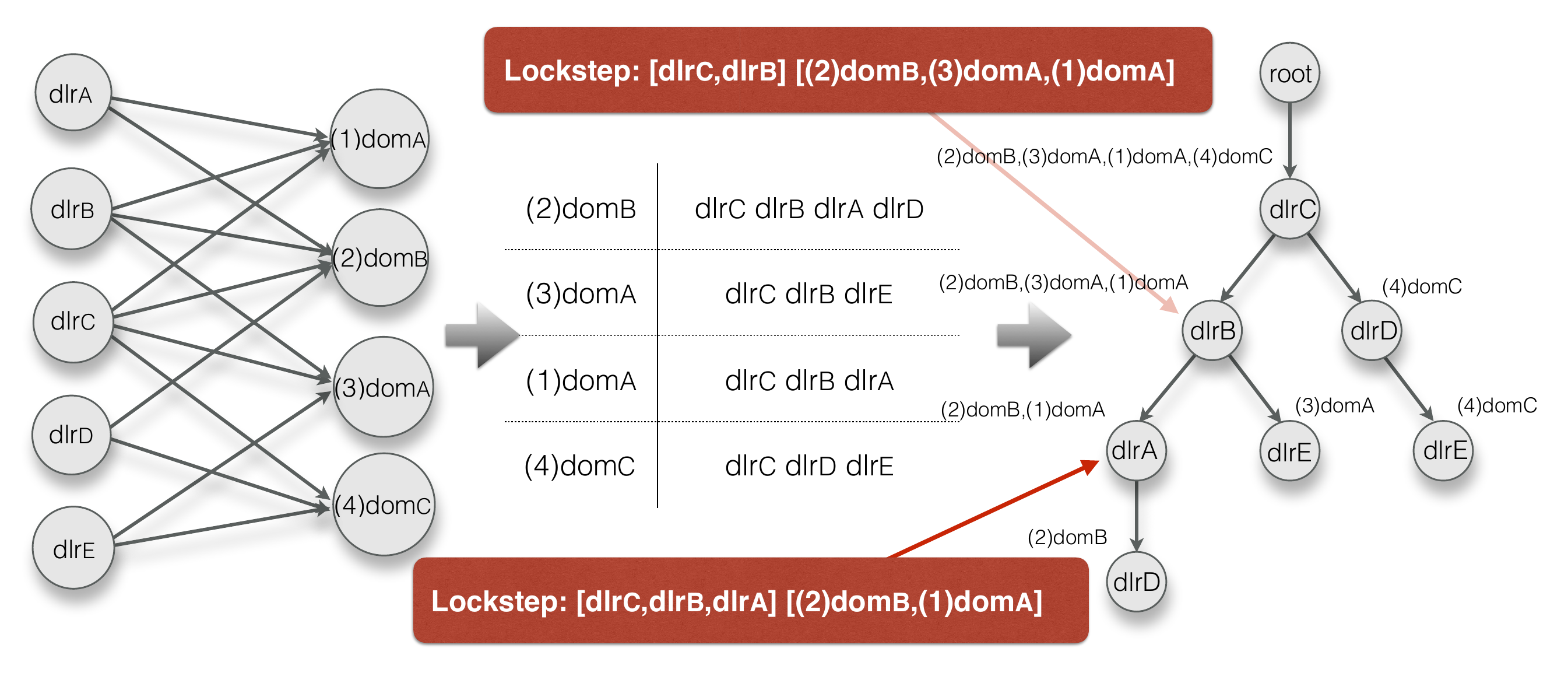}
\vspace{-5pt}
\caption{Example of FP tree construction: (a) Galaxy graph, (b) Sorted adjacency list, (c) FP tree.}
\vspace{-10pt}
\label{fig:fptree}
\end{figure}

Once the sorting is done, we start building the FP tree by 
creating a root $node_{fp}$ in the tree.
For each neighbor $u$ of $v$, we traverse the FP tree starting from the $root$ and check if $u$ is the child of the current $node_{fp}$. 
If this is the case, we set the current $node_{fp}$ as $u$ and append $v$ to its \emph{visited list}. 
Otherwise, we first add $u$ as the child of the current $node_{fp}$ and repeat the same process. 
We continue this process until we have checked all $node_{gg}$ $v's$ and their corresponding neighbors.
Figure~\ref{fig:fptree} illustrates the FP tree construction procedure given the \graph graph as input.  

Once the FP tree is constructed, we can traverse it to detect all the complete bicliques of the \graph graph. However, FP tree has some limitations : (a) FP tree does not return near-bicliques. (b) FP tree misses part of complete bicliques when overlap exists at the left hand nodes between a larger biclique and a smaller biclique. This results in the overlapped region being missed against the smaller biclique. We address how we handle these limitations in the next section.

\vspace{-3pt}
\subsection{Lockstep Detection}
\label{sec:lockstep-detection}
\vspace{-3pt}
After constructing the FP tree, we move to the lockstep detection phase. Each path downwards from the root to a $node_{fp}$ $A$ in the FP tree indicates a lockstep. The set of nodes along the path corresponds to the downloaders, and the visited list of $A$ corresponds to the domains in the lockstep. 
For example, in Figure~\ref{fig:fptree}, $dlr_{C}\rightarrow dlr_{B}\rightarrow dlr_{A}$, the resulting lockstep will be [($dom_{B}$, $dom_{A}$), ($dlr_{C}$, $dlr_{B}$, $dlr_{A}$)]. 
When identifying a lockstep, we remove the star id from the domain nodes; however, we store the star ids along with the lockstep, so that we do not lose the download events that resulted from the lockstep behavior. 
We can observe in Figure~\ref{fig:fptree} that some bicliques are not interesting; for example, when $A$ is a child of the root (e.g. $dlr_C$), we get a star centered on $A$, and when $A$ is a leaf (e.g. $dlr_E$), we get a star centered on the single domain from the visited list of $A$ (e.g. $dom_A$).
To avoid generating locksteps that are too small or that are a subset of a larger lockstep, we filter out the locksteps that satisfy the following conditions: (1) the number of downloaders or the domains are either below 3; or (2) $A$ has a child with same visited list. 

\topic{Partially missing locksteps} The FP tree captures most of the locksteps, however it misses the small locksteps that share part of the left hand elements with the larger lockstep. 
In Figure~\ref{fig:fptree}, we see that path $dlr_{C}\rightarrow dlr_{D}$ should have produced the lockstep of [($dom_{B}$, $dom_{C}$), ($dlr_{C}$, $dlr_{D}$)]. However, because $dlr_{C}$ and $dlr_{D}$ are the part of the longer path $dlr_{C}\rightarrow dlr_{B}\rightarrow dlr_{A}\rightarrow dlr_{D}$, $(2)dom_{B}$ fails to visit the corresponding path. We observe that this phenomenon occurs at the nodes that appear in multiple paths, such as $dlr_{D}$ and $dlr_{E}$ in our example.
We can recover the missing locksteps by maintaining different \emph{node versions}, for each path where the node appears, and by 
constructing a separate FP tree only on the stars that contain the node with multiple versions. 
To cover all the locksteps, we could do this recursively until there is no node with multiple versions in the FP tree. 
However, considering the overhead due to the recursive computation and the chance that the near-biclique algorithm would help recover some of the partially missing locksteps as
 explained in the next paragraph, we only apply the FP tree construction once on each nodes with multiple versions without recursion.

\topic{Near-bicliques} 
We aim to detect locksteps even in cases where some edges are missing from the \graph graph, e.g. the corresponding download events may have not been recorded for some reason. 
These missing edges could prevent some potential nodes to be added to the lockstep. 
Therefore we relax the lockstep definition, and search for subgraphs that include a fraction $\alpha \geq \alpha_{min}$ of the edges that would form a biclique. We set $\alpha_{min}$ to $0.8$ to accommodate for at most $1$ missing edge in the smallest lockstep.

There could be many possible missing edges. We reduce the search space by exploiting the fact that the adjacent nodes in the FP tree have higher connectivity than the other nodes, which implies that by introducing it into the lockstep will have fewer missing edges. 

We point to the end node $A$ of the path, which we want to extract the lockstep. We start by traversing the FP tree upwards, toward the root, until we reach a node $B$ that has a larger visited list. We also count the number of hops ($missing_{v}$)
required to reach $B$. We define the relative complement list as the difference between the visited list of $B$ and that of $A$. The relative complement list
will be added to the candidate list with $missing_{v}$ as an attribute. Next, we look at the children $A$. Each child will be added to the candidate list with the size of difference between its visited list and $A's$ visited list as the attribute $missing_{u}$. 

Once we get the candidate list, we sort it by the attribute in ascending order. Starting from the first node in the list, we add the node into the lockstep and calculate $\alpha$ which corresponds to the edge density within a lockstep. 
We stop when $\alpha$ 
drops below $\alpha_{min}$.
We observed that, in practice, 
this heuristic is good enough, as the adjacent nodes in the FP tree 
are more likely to be connected to the lockstep 
than the other nodes. 

\vspace{-3pt}
\subsection{Streaming Set-up}
\label{sec:streaming-setup}
\vspace{-3pt}
When using \system in a streaming setting, we ingest the download event data in real time. Instead of triggering our system for each single data stream, we run the system by processing incoming data as a batch within a fixed period $\Delta t$. Except for the difference in how the data comes in the system, the rest of the process is identical to that of the non-streaming setup. The star detection will search for new stars from the batch data; the new stars will be added to the \graph graph; the FP tree will be built from the \graph graph; and the lockstep detecion will find new locksteps.

%% file: campaigns_arxiv.tex
\vspace{-3pt}
\section{Silent distribution campaigns}
\label{sec:campaigns}
\vspace{-3pt}

In this section, we present a large scale empirical study of silent delivery campaigns. 
As discussed in Sections~\ref{sec:star-detection} and \ref{sec:bipartite-graph}, we can track two types of stars in the \graph graph:
multiple downloaders accessing a domain ($type_{dlr:dom}$) and vice versa ($type_{dom:dlr}$). 
These two star types result in different bicliques, and capture different download activities.
\label{assumption:different_lockstep_types} %
The difference derives from the fact that the central nodes in the stars may be duplicated in the \graph graph, when we add new stars that emerge in later time windows. 
The resulting locksteps reflect different distribution strategies. 
$type_{dlr:dom}$ account for downloaders that are more stable than the domains. 
Conversely, $type_{dom:dlr}$ identify distribution networks where domains are more stable.

For our empirical analysis, we set a narrow time window, to detect download events that are remotely triggered and do not experience delays. 
More generally, we should choose a shorter time window than the typical reaction time of domain blacklists during the observation period.
In consequence, we set the time window $\Delta t$ to 3 days, and we use a sliding window  $\delta t$ of 3 days.%
\footnote{During our observation period, domains delivering malware were blacklisted within 17 days on average~\protect\cite{DBLP:conf/raid/KuhrerRH14}.}

\label{sec:empirical-insights}

\begin{figure*}[t]
\vspace{-10pt}
\centering
\includegraphics[width=58mm]{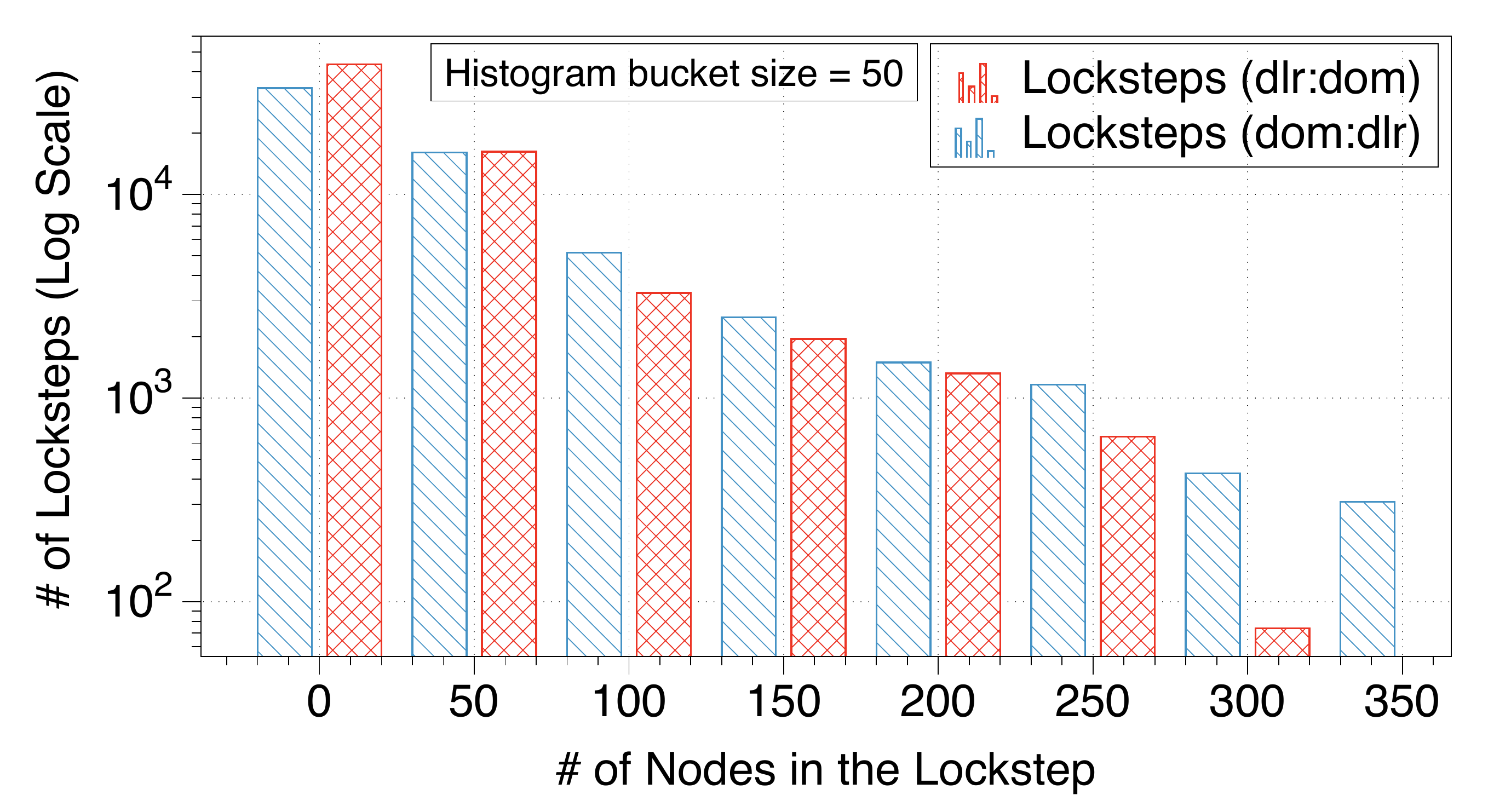}
\includegraphics[width=58mm]{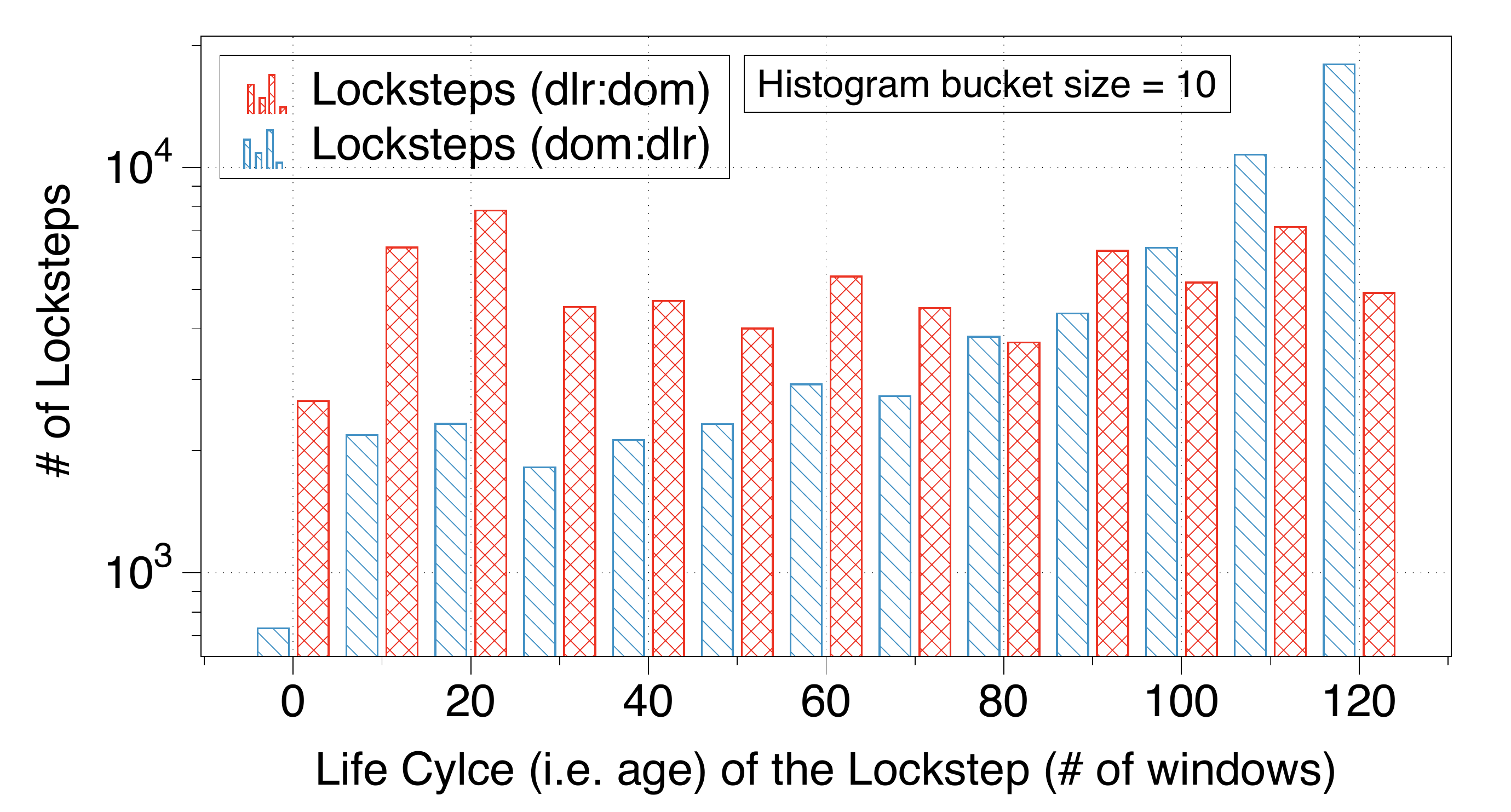}
\includegraphics[width=58mm]{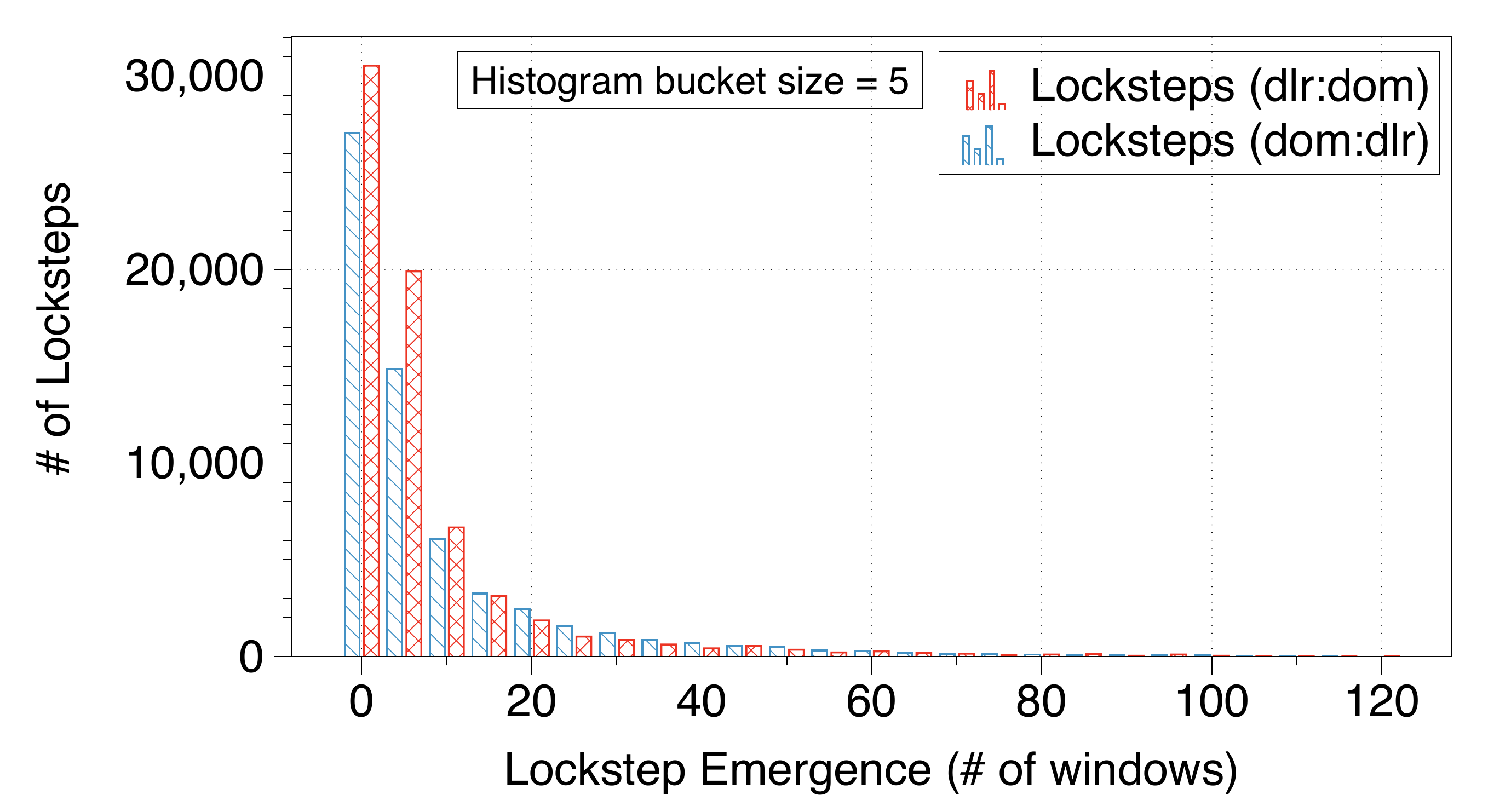}
\vspace{-5pt}
\caption{Distribution of lockstep properties: (a) Number of nodes, (b) Life span, (c) Lockstep emergence.} 
\vspace{-10pt}
\label{fig:locksteps_stat}
\end{figure*}

We identify \NumLockStepsVerDlr locksteps of type $type_{dlr:dom}$ and \NumLockStepsVerUrl locksteps of type $type_{dom:dlr}$ in our data. 
Figure~\ref{fig:locksteps_stat} illustrates the distributions for three properties of these locksteps: size, life span, and first detection time. 
The size of a lockstep corresponds to the number of nodes, considering both downloaders and domains in the lockstep. We deduplicate the central star nodes by removing the star IDs. Therefore, the number is counted on unique set of nodes in the lockstep.
For both types of locksteps, we observe that the number of nodes within each lockstep follows a long tail distribution i.e., there are many small locksteps and fewer large locksteps.
In Figure~\ref{fig:locksteps_stat}(b), we report the life span for each lockstep. 
There are long lived locksteps, enduring close to a year.
As our observation period spans only one year, there could be locksteps that live longer.  
To evaluate the opportunity for early discovery of malware delivery campaigns, 
for each lockstep we compute the delay until its first subset lockstep is formed,
which is the time difference between the addition of the second and the first star to the lockstep (Figure~\ref{fig:locksteps_stat}(c)).
We observe that, while the second star does not usually appear in the next time window, the locksteps nevertheless emerge quickly: 
both $type_{dlr:dom}$ and $type_{dom:dlr}$ take a median of 3 windows to form a lockstep.
However, we also observe some locksteps that emerge after a long delay.

\topic{Lockstep attribution} 
In general, it is challenging to identify precisely which organizations were controlling the download activities reflected in the locksteps we detect, as the domains may no longer be registered and the downloaders may no longer be active. 
However, we aim to make a coarse grained distinction among the distribution campaigns for malware, PUP and benign software, to compare their properties and to assess their overlaps. 
To do so, we observe that 38.2\% (3479 out of 9103) of the downloaders involved in locksteps are digitally signed, with valid X.509 certificates.
We first 
analyze these signatures to determine the most frequent publisher in a lockstep. 
We consider that a publisher is the \textit{representative publisher (rep-pub)} of the lockstep, if 
it accounts for more than 50\% of the signed downloaders in the lockstep. 
If we cannot identify a representative publisher, we set the lockstep's \textit{rep-pub} to \verb!mixed!. 
In this manner, we identify 335 \textit{rep-pubs}. 
We investigate the top 50 \textit{rep-pubs} from each lockstep type and we manually categorize them into 6 different groups: 
potentially unwanted programs (PUP)~\cite{kotzias2015pup},
pay-per-install (PPI)~\cite{caballero2011measuring},
benign (BN),
other,
mixed,
and unknown (UK). The first 4 groups inherit the label of the \textit{rep-pub}, determined as discussed in Section~\ref{sec:groundtruth}. 
We place the \verb!mixed! \textit{rep-pubs} 
in a separate group.
In some cases we cannot identify the real publisher behind the lockstep, as the downloader is an archive extractor (Winzip). 
These correspond to the unknown group.  
Table~\ref{table:lockstep_groups} describes the distribution of these lockstep groups. While we are able to label some locksteps in this manner, we observe that most locksteps involve downloaders that are difficult to place in a specific category, 
as many locksteps have mixed rep-pubs. %

\begin{table}[t]
\centering
\caption{Lockstep group statistics.}
\label{table:lockstep_groups}
\scalebox{0.9}{
\begin{tabular}{l|l|l|}
\cline{2-3}
\multicolumn{1}{c|}{}       & \begin{tabular}[c]{@{}l@{}}$type_{dlr:dom}$\\ (MDL/PDL/BDL/UDL)\end{tabular} & \begin{tabular}[c]{@{}l@{}}$type_{dom:dlr}$\\ (MDL/PDL/BDL/UDL)\end{tabular} \\ \hline
\multicolumn{1}{|l|}{PUP}   & \begin{tabular}[c]{@{}l@{}}27,522\\ (26,764/501/109/148)\end{tabular}             & \begin{tabular}[c]{@{}l@{}}13,117\\ (11,902/1,202/6/7)\end{tabular}             \\ \hline
\multicolumn{1}{|l|}{PPI}   & \begin{tabular}[c]{@{}l@{}}2,639\\ (2,137/498/4/0)\end{tabular}             & \begin{tabular}[c]{@{}l@{}}1,496\\ (1,164/332/0/0)\end{tabular}             \\ \hline
\multicolumn{1}{|l|}{BN}    & \begin{tabular}[c]{@{}l@{}}3,939\\ (1,749/888/597/705)\end{tabular}       & \begin{tabular}[c]{@{}l@{}}2,021\\ (1,152/840/7/22)\end{tabular}         \\ \hline
\multicolumn{1}{|l|}{Other} & \begin{tabular}[c]{@{}l@{}}9,203\\ (8,041/1,053/58/51)\end{tabular}       & \begin{tabular}[c]{@{}l@{}}5,092\\ (3,479/1,580/8/25)\end{tabular}              \\ \hline
\multicolumn{1}{|l|}{Mixed} & \begin{tabular}[c]{@{}l@{}}20,766\\ (14,085/4,069/2,255/357)\end{tabular}       & \begin{tabular}[c]{@{}l@{}}36,594\\ (32,576/2,479/1,449/90)\end{tabular}       \\ \hline
\multicolumn{1}{|l|}{UK}    & \begin{tabular}[c]{@{}l@{}}86\\ (86/0/0/0)\end{tabular}            & \begin{tabular}[c]{@{}l@{}}835\\ (808/27/0/0)\end{tabular}            \\ \hline
\end{tabular}}
\vspace{-10pt}
\end{table}

We therefore perform a second labeling step, 
based on the payloads that the locksteps distribute. 
We distinguish between malware and PUP payloads with the method described in Section~\ref{malware-pup-distinction}.
The labeling is conducted in two steps.
First, the downloaders are labeled by the payloads they distribute within the lockstep. 
We say a downloader is \textit{malware downloader (MD)}, 
if it distributes at least one malware.%
\footnote{This is an aggressive labeling policy, as even benign downloaders may be tricked into downloading malware occasionally. However, this labeling produces a conservative estimate of our false positive rate (as discussed in Section~\ref{sec:problem}, we do not aim to measure false negatives).}
In a similar fashion, we label a downloader as \textit{PUP downloader (PD)} 
if it downloads PUP payloads but no malware. 
A downloader is labeled as \textit{Benign downloader (BD)} 
if it downloads a benign payload but no suspicious (malware, PUP) download. 
The rest are placed as \textit{unknown downloader (UD)}. 
As the next step, we label the locksteps. 
The locksteps that include at least one MD are labeled as \textit{malware downloader lockstep (MDL)}. 
Similarly, we label a lockstep as \textit{PUP downloader lockstep (PDL)} if it contains PDs but no MDs. 
We label the locksteps with no suspicious (MD, PD) downloader as \textit{unknown downloader lockstep (UDL)}. 
We note that, as malware families sometimes evade detection for extended periods of time, not every UDLs correspond to benign download activities. 
Therefore, we try to identify the \textit{benign downloader locksteps (BDL)} among the UDLs. Similar to the definition of MDLs and PDLs, the BDL should contain at least one BD. 
\begin{table}[t]
\centering
\caption{Lockstep label statistics}
\label{table:lockstep_label_stat}
\begin{tabular}{l|l|l|}
\cline{2-3}
\multicolumn{1}{c|}{}           & $type_{dlr:dom}$ & $type_{dom:dlr}$ \\ \hline
\multicolumn{1}{|l|}{MDL} & 54,497 (81.22\%) & 51,831 (85.81\%) \\ \hline
\multicolumn{1}{|l|}{PDL} & 7,800 (11.63\%) & 6,901 (11.43\%) \\ \hline
\multicolumn{1}{|l|}{BDL}    & 3,231 (4.82\%)   & 1,500 (2.48\%)      \\ \hline
\multicolumn{1}{|l|}{UDL} & 1,566 (2.33\%)  & 169 (0.28\%)  \\ \hline
\end{tabular}
\vspace{-10pt}
\end{table}
We present the result of the labeling in Table~\ref{table:lockstep_label_stat}. For both lockstep types, MDL occupy more than 80\% of the total number of locksteps while benign are 4.82\% and 2.48\% for $type_{dlr:dom}$ and $type_{dom:dlr}$, respectively. 
Our higher success rate in labeling with payloads, compared to labeling only with downloaders, reflects our community's focus on detecting and labeling malware, rather than on understanding the client-side distribution infrastructure. 

\topic{Identifying campaigns}
As discussed in Section~\ref{sec:problem}, we separate the campaigns within the lockstep by $n\delta t$. 
By setting $n=3$,
we identify 1,292,141/71,424/27,145/6,233 campaigns corresponding to MDL/PDL/BDL/UDL. On average there are 12.2/4.9/5.7/3.6 campaigns per lockstep for MDL/PDL/BDL.

\vspace{-3pt}
\subsection{Relationships among representative publishers}
\label{sec:business-relationships}
\begin{figure*}[t]
\vspace{-10pt}
\centering
\includegraphics[width=100mm]{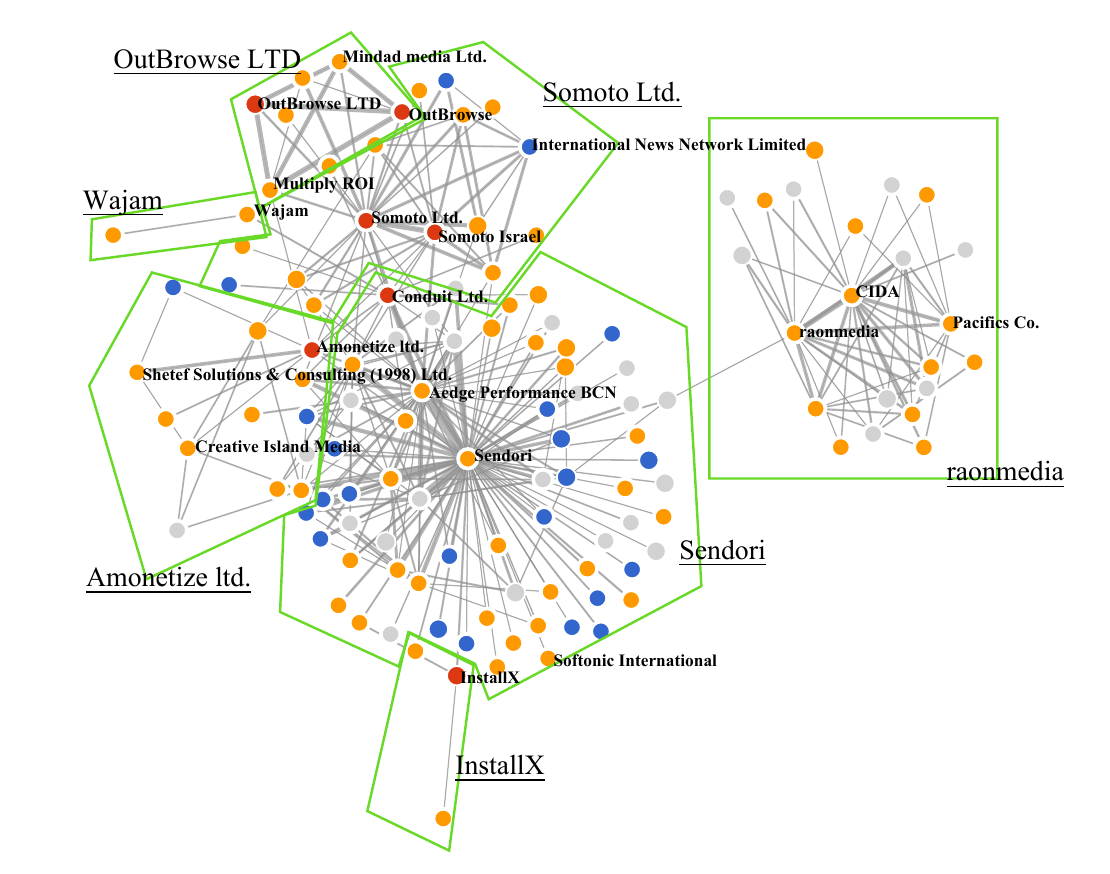}
\includegraphics[width=60mm]{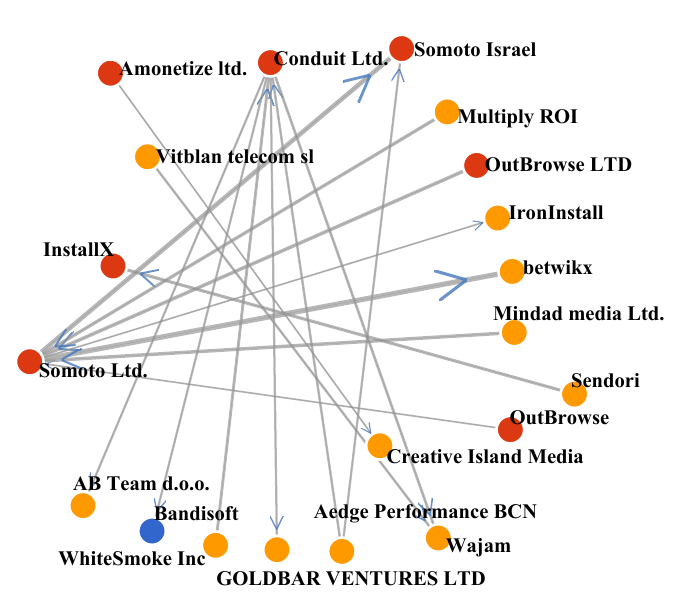}
\vspace{-5pt}
\caption{Business relationship: (a) Both partner and neighbor, (b) partner relationship for PPIs. (node color red/orange/blue/gray corresponds to PPI/PUP/benign/other).}
\vspace{-10pt}
\label{fig:business_rel}
\end{figure*}

\vspace{-3pt}
The locksteps allow us to determine the business relationships 
between rep-pubs and payloads and 
among groups of rep-pubs.
We focus on PPI and PUP providers, which distribute other executables intentionally. 
We collect PUP and PPIs from the top 10 rep-pubs 
with a high percentage of MDLs within their locksteps.
Each of these rep-pubs conducted at least 40 campaigns. 
We also include the known PPI providers \verb!Amonetize Ltd.!, \verb!Conduit Ltd.! and \verb!OutBrowse LTD! to this list.
We investigate which publishers appear frequently together in lockstep with these 13 rep-pubs.
As the downloaders signed by these publishers simultaneously utilize the same server side infrastructure, this likely reflects a relationship among the corresponding distribution networks. 
We also determine whether one of these downloaders was itself downloaded by 
one of the the downloaders in the lockstep, which suggests a closer business connection.
We therefore term such relationship between the publishers as \emph{partner}.
For example, we observe such partnership relations among some PPI providers, e.g.
\verb!Outbrowse Ltd.! that frequently delivers downloaders from \verb!Somoto Ltd.!.
Additional frequent partners of \verb!Somoto Ltd.! include \verb!Mindad media Ltd.!, \verb!IronInstall!, \verb!betwlkx!, and \verb!Multiply ROI!, which suggests a stable business relationship with these organizations.\footnote{Several of these publishers attended the 2014 Affiliate Summit in Las Vegas (\url{http://affiliatesummit.com/}).} 
The cases where we cannot establish a \emph{downloaded-by} relationship among the downloaders in the lockstep may point to an organization that uses multiple code signing certificates to evade attribution or to relationships with a common third party. 
We term such relationship as \emph{neighbor}.
We illustrate some of these business relationships 
in Figure~\ref{fig:business_rel}. The nodes are the publishers and the edge between publishers indicate a business relationship, either \emph{partner} or \emph{neighbor}. The thickness of the edge indicates the frequency of that relationship.
To further illuminate this ecosystem, we employ a community detection algorithm \cite{fastgreedy} to the graph illustrated in Figure~\ref{fig:business_rel}(a).
This algorithm identifies 7 communities. 
Within each community, we determine the rep-pub with the highest betweenness centrality~\cite{freeman1977}, which is the number of shortest paths between any two nodes that pass through the rep-pub. 
This graph centrality measure singles out a node that likely acts a bridge between other nodes from each community. 

\squishlisttwo
\item {\em Community \#1: OutBrowse.} This community represents the advertisers or the affiliates of 
the Outbrowse PPI. 
The PUPs \verb!Multiply ROI! and \verb!Mindad media Ltd.! are frequently in lockstep with the rep-pub. 
The other publishers in this community represent variants of the rep-pub's certificate: \verb!OutBrowse LTD! and \verb!OutBrowse!.
\item {\em Community \#2: Somoto.} This community belongs to Somoto, which is also a PPI provider. Beside Somoto's certificates (\verb!Somoto Ltd.! and \verb!Somoto Israel!), this community includes 12 other publishers. 
\texttt{International News Network Limited}, a known PUP distributor, is tightly connected with the publishers in this community,
suggesting a close relationship.
\item {\em Community \#3: raonmedia.} 22 publishers belong to this community. Three PUP publishers including \verb!raonmedia!, \verb!Pacifics Co.!, and \verb!CIDA! showed high centrality in this community. All three publishers were located in Busan, Korea and the certificates were issued by Thawte, Inc., which suggests these publishers could belong to the same group.  
\item {\em Community \#4: Sendori.} Although we see PPI \verb!Conduit Ltd.! within the community, PUP \verb!Sendori! has a higher centrality. At 77 publishers, this is the largest community. \verb!Sendori! was is tightly connected to most of the publishers within the group, which reflects an aggressive distribution strategy of this PUP.
\item {\em Community \#5: Amonetize.} This group represents \verb!Amonetize Ltd.! and several PUPs. In particular, \verb!Shetef! \verb!Solutions &! \verb!Consulting (1998) Ltd.! is known to be the advertiser\footnote{\url{https://www.reasoncoresecurity.com/signer-shetef-solutions-consulting-1998-ltd-40812da0f7cb2ecd495} 5fd76e0a6c493.aspx} of Amonetize. 
\item {\em Communities \#6 \& \#7.} These communities 
are small %
and include the \verb!InstallX! PPI %
and the \verb!Wajam! PUP. %
\squishend

These results suggest that the partner and neighbor relationships can 
expose organizations that utilize distinct code signing certificates for different activities, e.g. PPI and PUP. 
Additionally, the graph communities capture close relationships among the publishers, such as delivery networks that rent the server-side infrastructure from a third party or publishers that engage in aggressive distribution campaigns using multiple providers. 
The graph also includes
instant messengers and file sharing software, which are likely involved in locksteps resulting from spam campaigns.

\vspace{-3pt}
\subsection{Malware and PUP delivery ecosystems}
\label{sec:malware-pup-overlap}
\vspace{-3pt}
Downloaders that appear in locksteps with different labels provide the opportunity to analyze the overlap of different software distribution ecosystems. 
36.7\% of the downloaders (3,345 out of 9,103) are present in both MDLs and PDLs. 
These downloaders are associated with 7,635 and 6,886 of $type_{dlr:dom}$ and $type_{dom:dlr}$ PDLs, which account for 97.8\% and 99.8\% of all the PDLs. 
100 of these downloaders dropped payloads known to be malicious, while the other ones downloaded other files in lockstep with the malware droppers. %
The PUP publishers from Figure~\ref{fig:business_rel} distributed 
13 trojan families, including \verb!vundo!, \verb!pasta!, \verb!symmi!, \verb!crone!, \verb!pahador!, \verb!pecompact!, \verb!scar!, \verb!dapato!, \verb!renum!, \verb!jorik!, \verb!fareit!, \verb!llac!, and \verb!kazy!. 
We also observed generic trojans, \verb!induc! (virus), \verb!zeroaccess! (botnet), \verb!onescan! (fakeAV), \verb!pincav! (keystroke logger), \verb!dnschanger!, \verb!startpage!, and several worms delivered through these publishers. 

To further illuminate the connection of the malware and PUP delivery ecosystems, we compare the publishers from our locksteps to the ones from the Malsign blacklist of certificates used to sign PUP and malware payloads~\cite{kotzias2015pup}.
In this way, we identify 1,926 downloaders signed by 212 publishers from malsign, which were involved in 70,984 and 5,468 of MDLs and PDLs respectively.
This suggests that many publishers thought to belong to the PUP category are also involved in malware delivery. 
Considering that many of the unknown files in our data set may be malware samples (83\% of our payloads were never submitted to VirusTotal), the number of MDLs is likely higher. %

These results contradict two recent studies~\cite{Thomas16:CommercialPPI, Kotzias16:PUP-PPI}, which did not find a substantial overlap between the malware and PUP delivery ecosystems. 
The key distinction is that these studies analyzed \emph{direct download relationships} between publisher pairs, while lockstep detection allows us to identify \emph{indirect relationships}, through the neighbor links discussed in Section~\ref{sec:business-relationships}.
These indirect links can overcome evasive strategies such as certificate polymorphism or utilizing unsigned downloaders for malicious payloads. 
In particular, in Somoto's locksteps, 90.6\% of the downloaders, on average, are either unsigned or have invalid certificates. 
We also observe several PUPs with over 50\% ratio, including \verb!Strongvault Online Storage LLC!, \verb!Save Valet!, and \verb!LLC Mail.Ru!.
Variations in experimental methods may further explain the different results. 
Thomas et al.~\cite{Thomas16:CommercialPPI} milk PPI downloaders on hosts located in the US, while our data set includes hosts from 72 countries. 
Geographical targeting has been reported previously for PPI providers~\cite{caballero2011measuring}.
Additionally, their data set covers a different observation period. 
In contrast, Kotzias et al.~\cite{Kotzias16:PUP-PPI} analyze data from WINE from a time span that largely overlaps with our observation period. 
However, they focus on 70 malware families, excluding for instance trojans that received generic labels from anti-virus vendors.

\vspace{-3pt}
\subsection{Properties of MDLs}
\label{sec:mdl-properties}
\vspace{-3pt}

We identify a total of 54,497 and 51,831 locksteps of $type_{dlr:dom}$ and $type_{dom:dlr}$, respectively, that download at least one malware. These MDLs come from 246 and 169 \textit{rep-pub} each.
In addition to the PPI and PUP delivery vectors discussed above, we observed that malware is sometimes distributed through compromised software updates.
We identified a 
malware distribution campaign involving the \verb!KMP Media Co.!, which is a legitimate media player.  
The campaign distributed trojan \verb!dofoil!. 
The version of the media player involved in the MDL is 3.6.0.87, which is known to have a stack overflow vulnerability\footnote{\url{https://www.krcert.or.kr/data/secNoticeView.do?bulletin_writing_sequence=2147&queryString=cGFnZT0xJnNvcnRfY29kZT0mc2VhcmNoX3NvcnQ9a2V5d29y} ZCZzZWFyY2hfd29yZD1pb3M=} that was exploited in the wild.\footnote{\url{https://www.virustotal.com/en/file/ff49e145515bdecbca61b7d9} 7439959be5b04b1c29d77a0e8c42a1c1bed42aa8}
Additionally, while experimenting with larger values for $\Delta t$, we observed a Hewlett-Packard software updater deliver the \verb!hexzone! ransomware.\footnote{We did not find evidence that HP's code signing certificate was compromised; it is more likely that the malware was able to infect the server-side infrastructure involved in software updates. This is consistent with prior reports of a trojan that was signed by HP after it infected the company's systems, but without having compromised any certificates~\cite{Krebs15:SignedMalware}.}

\begin{figure}[t]
\vspace{-5pt}
\centering
\includegraphics[width=65mm]{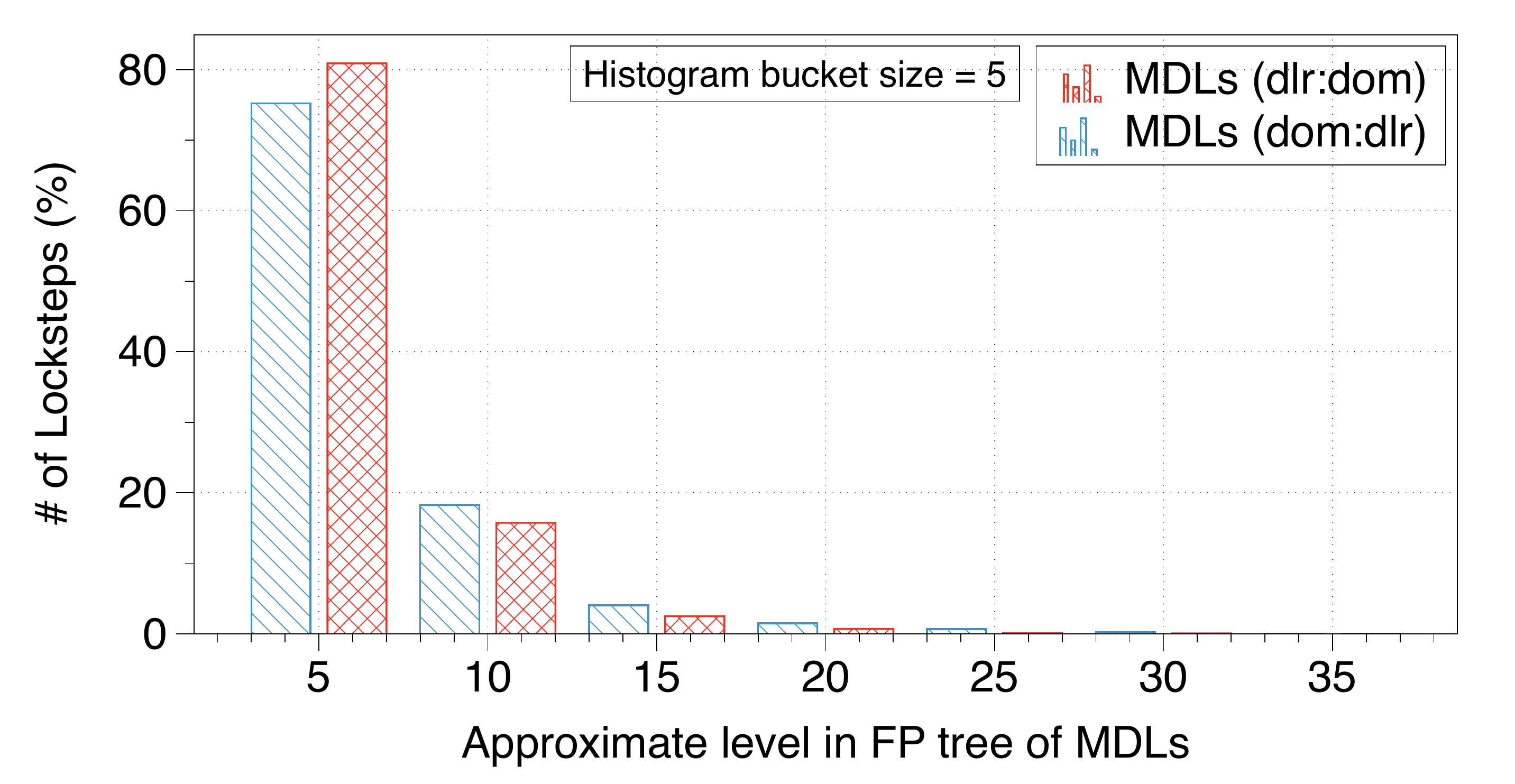}
\vspace{-5pt}
\caption{Approximate FP Tree Level of the MDLs.}
\vspace{-10pt}
\label{fig:fptree_level_mal_lck}
\end{figure}

We observe several features that distinguish MDLs from other locksteps.
Figure~\ref{fig:fptree_level_mal_lck} illustrates the approximate FP tree level where the MDLs reside. 
As each node in the FP tree corresponds to a lockstep, the typical level of MDLs indicates the region of the FP tree where we are most likely to find evidence of malware distribution. 
This is an approximation, as we may add or subtract a level when computing near-bicliques, as described in Section~\ref{sec:lockstep-detection}.
The median FP tree level where locksteps reside is 5, for both $type_{dlr:dom}$ and $type_{dom:dlr}$. 
In other words, the median number of downloaders in a MDL is 5. 
This is relatively close to the root of the FP tree, as malware delivery networks rely on only a few downloaders within a time window. 
This observation helps us improve the performance of \system in streaming mode, as discussed in Section~\ref{sec:evaluation}.

\begin{figure*}[t]
\vspace{-10pt}
\includegraphics[width=58mm]{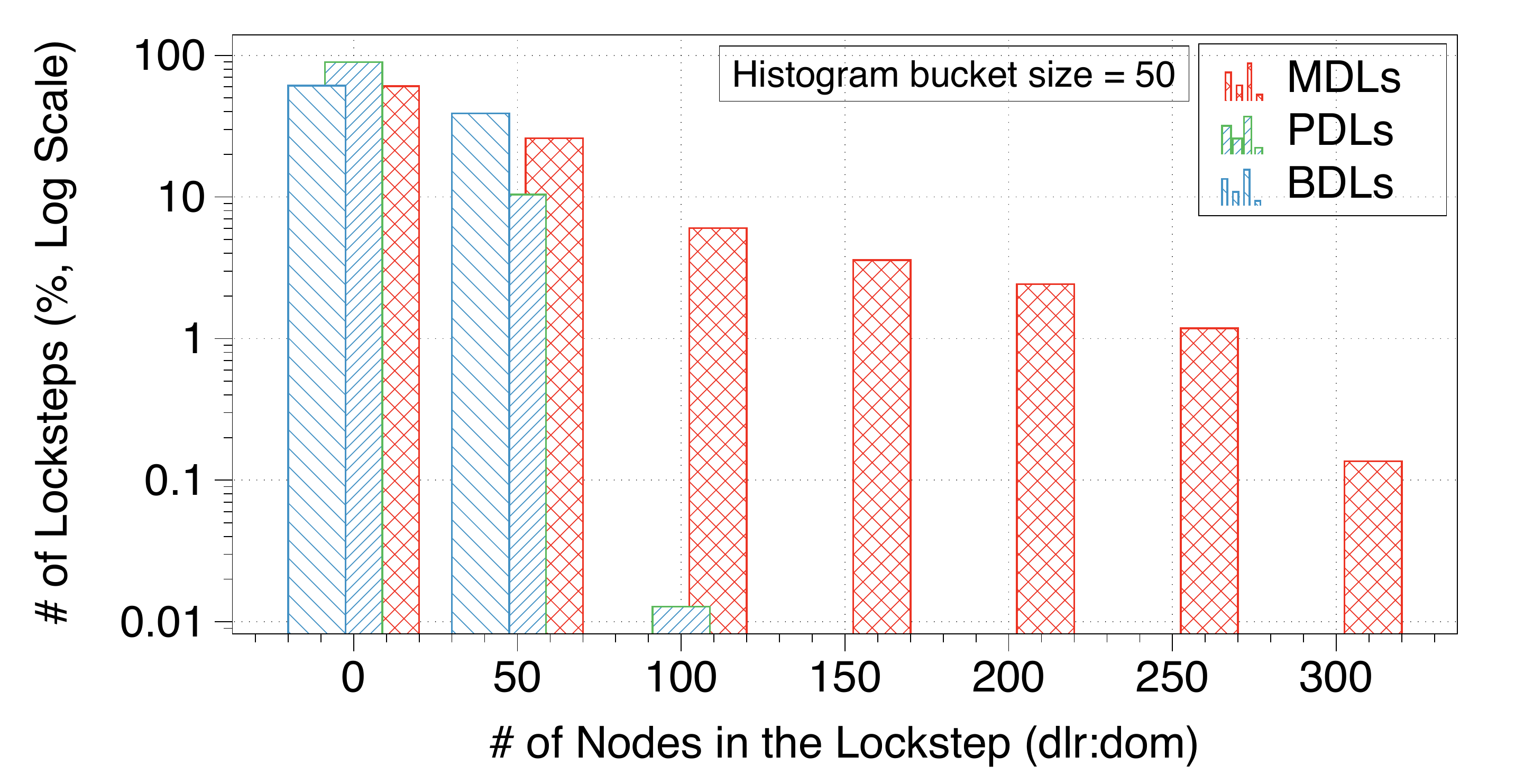}
\includegraphics[width=58mm]{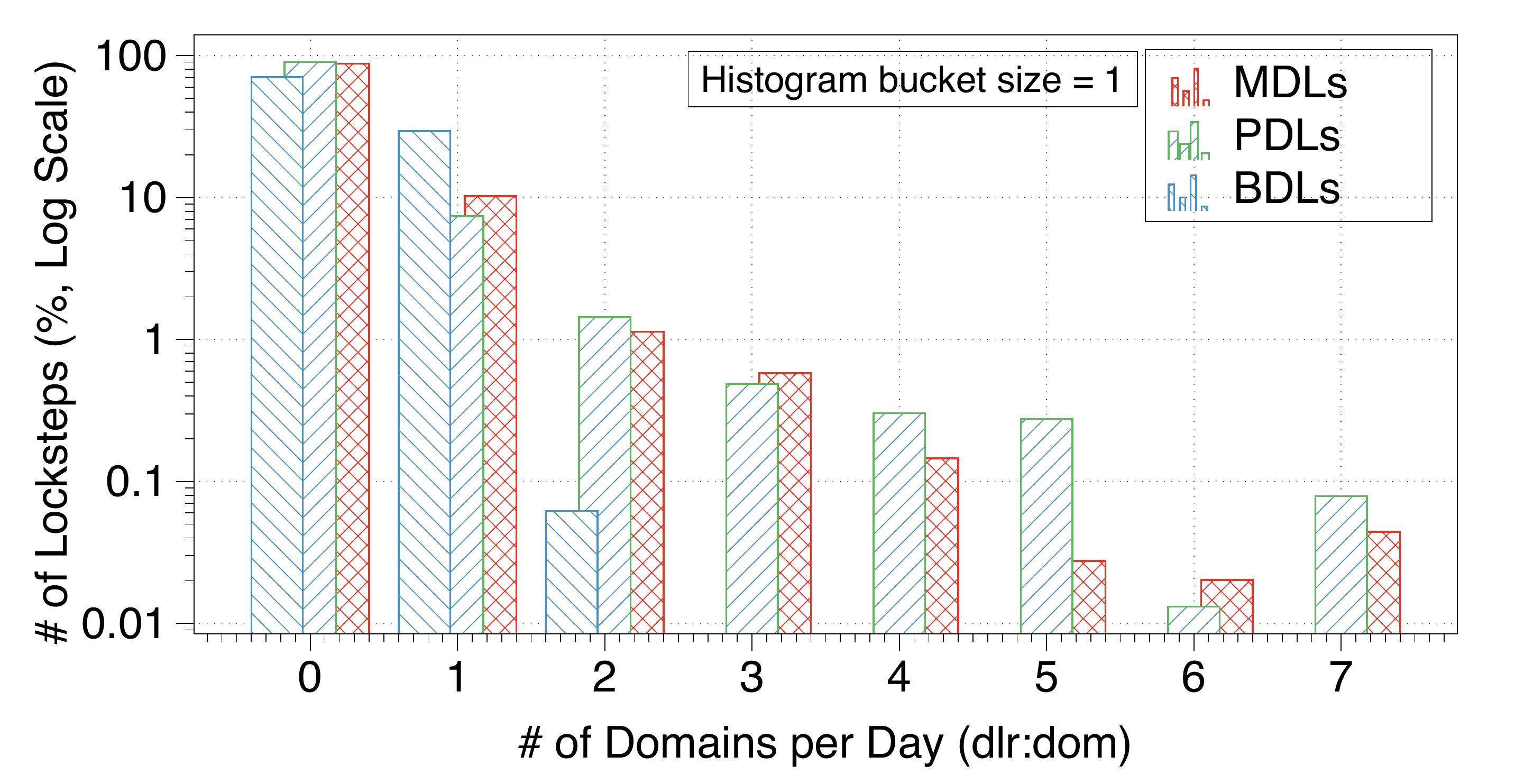}
\includegraphics[width=58mm]{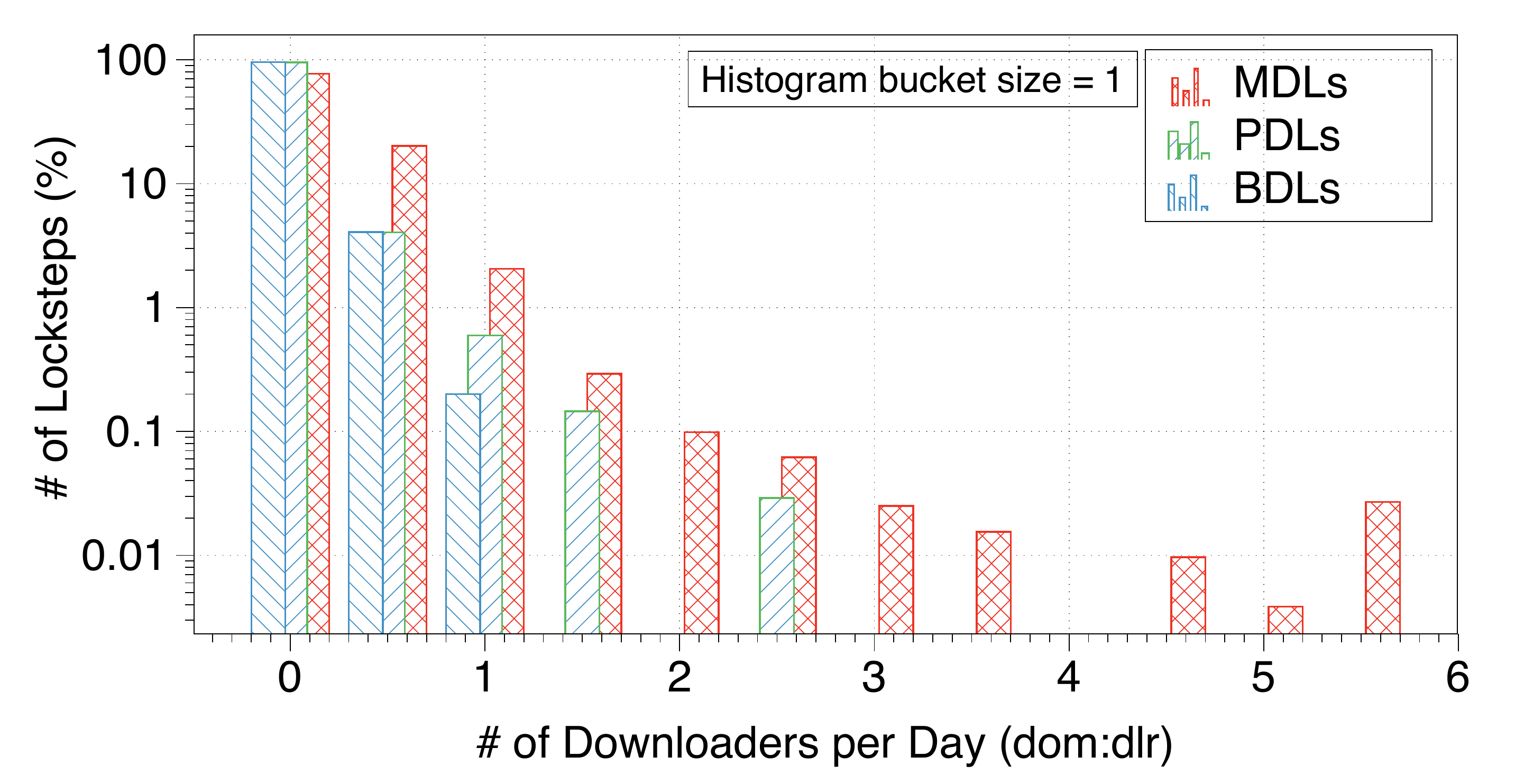}
\vspace{-5pt}
\caption{MDL properties:
(a) Distribution of the number of nodes in lockstep,
(b) Distribution of the number of domains per day ($type_{dlr:dom}$),
(c) Distribution of the number of downloaders per day ($type_{dom:dlr}$).
}
\vspace{-10pt}
\label{fig:visualexploration}
\end{figure*}

We also observe that MDLs tend to have a large number of nodes, as illustrated in Figure~\ref{fig:visualexploration}(a). 
In contrast, although we see a few large PDLs, 
around 90\% PDLs have fewer than 25 nodes and over 90\% BDLs have fewer than 75 nodes. 
Figure~\ref{fig:visualexploration}(b) illustrates the number of domains per day for each locksteps. We observed MDLs showing aggressive domain churn (more than 7 domains per day).
Figure~\ref{fig:visualexploration}(c) illustrates the number of downloaders per day for each lockstep. On average, a new downloader appears for every 5.8/16.7/11.8 days for MDLs/PDLs/BDLs respectively. We also observed MDLs showing aggressive downloader repacking (more than 5 downloaders per day).

\vspace{-3pt}
\section{Detection performance}
\label{sec:detection-performance}
\vspace{-3pt}
While the previous section provides empirical insights into silent distribution campaigns, we now evaluate the effectiveness of \system as a detection system. 
We aim to detect suspicious activity, such as malware and PUP dissemination campaigns. 
This information can be used in several ways. 
The downloaders and domains caught in locksteps can help prioritize further analysis, e.g. to attribute the campaigns to publishers as we demonstrate in Section~\ref{sec:empirical-insights}. 
It could be combined with other techniques (e.g. DNS reputation systems~\cite{antonakakis2010, bilge2014}) to detect a specific form of abuse (e.g. botnet activity).
An enterprise may also block all downloads initiated remotely by unknown organizations; in this case, a few trusted publishers could be added to our initial whitelist.

We use the locksteps labeled in Section~\ref{sec:empirical-insights} to validate our system: an MDL or PDL detection represents a \emph{true positive}, while a BDL detection is a \emph{false positive}.
For the true positives, we compute the detection lead time, compared with the anti-virus products invoked by VirusTotal (for downloaders) and with three malware blacklists (for domains). 
We also analyze the causes of false positive detections.
As we lack ground truth about malware distribution campaigns, we cannot estimate the false negative rate.

\topic{Experimental settings}
We evaluate \system in offline mode, and we build on our empirical insights to select the appropriate configuration parameters.
We set $\Delta t = \delta t = 3$~days, to capture locksteps with a high domain churn. 

\vspace{-3pt}
\subsection{Malware and PUP detection}
\label{sec:malware-pup-detection-performance}
\vspace{-3pt}
\topic{Detection performance}
Table~\ref{table:lockstep_label_stat} lists the numbers of locksteps from each category. 
Overall, the benign locksteps (BDLs) represent 4.82\% and 2.48\% of $type_{dlr:dom}$ and $type_{dom:dlr}$ locksteps, respectively.
We observe the highest fraction of BDLs among the mixed locksteps of $type_{dlr:dom}$, perhaps because malware and PUP creators utilize dedicated malicious infrastructures as well as generic downloaders, which may also distribute benign software. 
In contrast, PPI rep-pubs do not generate \emph{any} BDL of $type_{dom:dlr}$ and only 4 BDLs of $type_{dlr:dom}$.
Overall, the suspicious locksteps (MDL or PDL) account for 92.85\% and 97.24\% of all locksteps of $type_{dlr:dom}$ and $type_{dom:dlr}$, respectively.

\begin{figure}[t]
\centering
\vspace{-5pt}
\includegraphics[width=65mm]{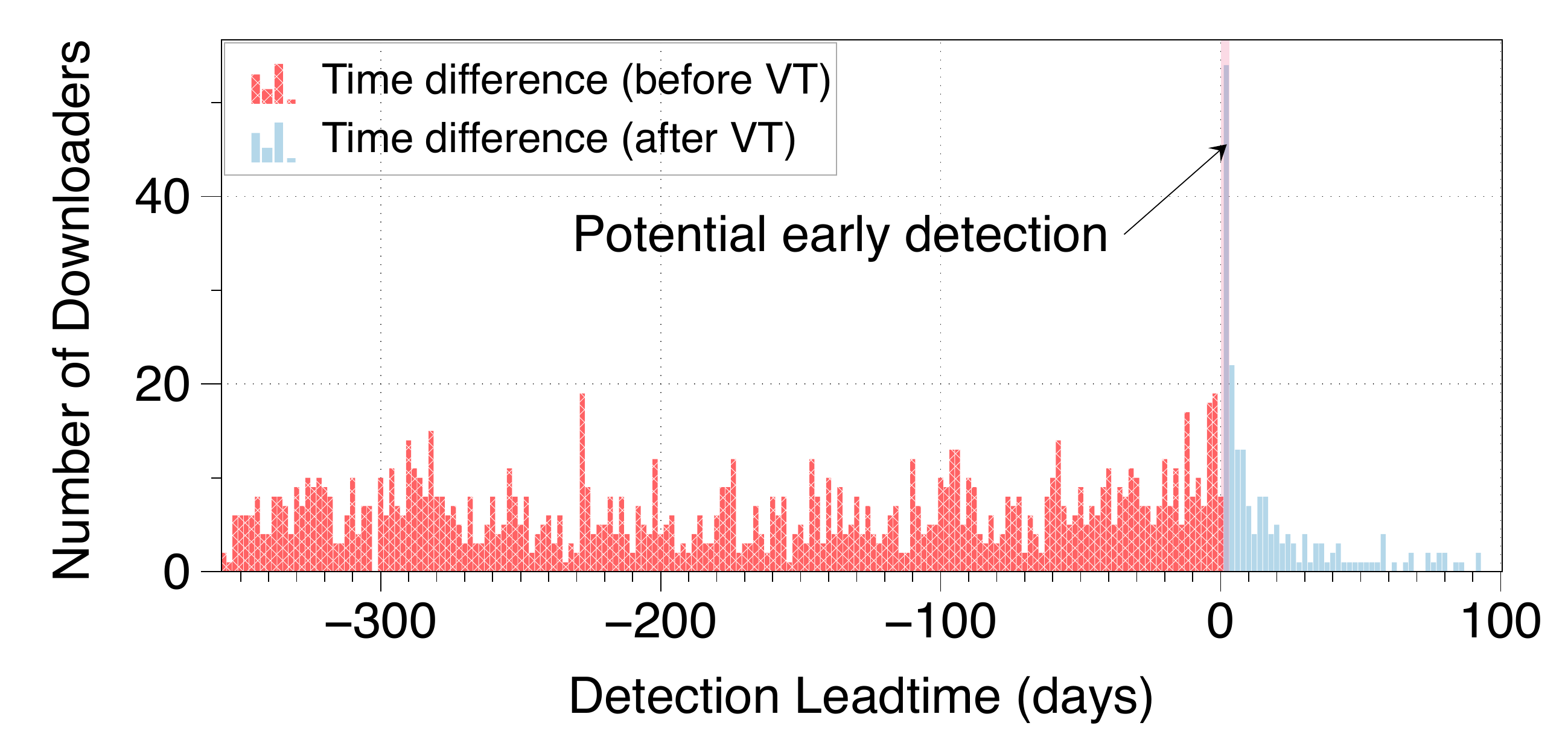}
\vspace{-5pt}
\caption{Detection lead time for MD/PDs.}
\vspace{-5pt}
\label{fig:detection_lead_dlr}
\end{figure}

\topic{Detection lead time}
As \system is content-agnostic (i.e. it does not analyze the downloader binaries or the Web content served by the URLs contacted), we evaluate how early we can detect suspicious downloaders or domains that are previously unknown. 
We consider the downloaders submitted to VirusTotal in 2013 that have at least one detection record.
We compare the time when \system is able to detect these downloaders to the time of their first submission to VirusTotal. 
Because a downloader detected by \system is active in the wild, and because VirusTotal invokes up to 54 AV products with updated virus definitions, we consider that a detection lead time illustrates the opportunity to identify previously unknown droppers. 
As explained in Section~\ref{sec:empirical-insights}, a lockstep emerges 
at the time when the second star is formed; we estimate the detection time of a downloader as the earliest detection timestamp among the locksteps that contain it.
Figure~\ref{fig:detection_lead_dlr} illustrates this comparison. %
The negative range represents a detection lead time, and the positive range corresponds to detection lag. 
We observe 1182 downloaders detected early and 213 downloaders detected late. 
The median detection lead time is 165 days. 
Among the late detections, 69 of the downloaders are detected $<$3 days late, which suggests that they may detected early 
with a shorter $\Delta t$. 
In contrast, the detection lead time seems uniformly distributed, suggesting that \system can detect both recent distribution campaigns as well as campaigns that have been operating for a while.

We also collect URLs blacklisted in 2013 from three publicly available sources \footnote{\url{https://www.malwaredomainlist.com}}\footnote{\url{http://www.malwaredomains.com}}\footnote{\url{https://www.phishtank.com}}.
These URLs correspond to 394 unique domains, of which 29 were present in our dataset. 
Among these 29 domains, 14 domains were caught in locksteps; the other 15 domains may represent false negatives, or they may correspond to malware dissemination techniques other than silent delivery campaigns. 
As for downloaders, we estimate the detection lead time for these 14 domains 
by comparing the lockstep detection dates with the blacklisting dates. 
Except for one domain that is detected 36 days later, 13 out of 14 domains are detected early, with an median lead detection time of 196 days.

\topic{False positive analysis}
We identified 80 publishers that are involved in forming the BDLs. The top 50 publishers account for 50\% of the downloaders in BDLs. 
22 of these publishers are benign, but they are absent from NSRL so they were not included in our whitelist.
These were mainly non-US publishers (\verb!ESTsoft Corp.!, \verb!AhnLab!, and \verb!NHN corp.!), which are not covered by NSRL, and benign publishers with multiple code-signing certificates (\verb!Skype Limited! is listed in NSRL but \verb!Skype Software Sarl! is not).
These BDLs could have been avoided with a more comprehensive whitelist.
Additionally, 17 publishers are labeled as Other, 12 as PUP, and 1 as PPI. 
We suspect that many of their locksteps deliver undetected malware or PUPs, as VirusTotal reports existed for only $\approx$17\% of payloads.

\vspace{-3pt}
\subsection{Comparison with alternative techniques}
\label{sec:prior-techniques-comparison}
\vspace{-3pt}
We compare our lockstep detection algorithm with two alternative techniques for detecting malicious campaigns: 
community detection algorithms \cite{fastgreedy}, \cite{multilevel}, which have been explored extensively in the context of graph mining,
and prior algorithm for detecting lockstep behavior \cite{copycatch}.

\topic{Community Detection}
To compare lockstep detection and community detection algorithms, we construct a $type_{dlr:dom}$ bipartite graph with all the download events.
We employ 2 popular community detection algorithms \cite{fastgreedy, multilevel} based on optimizing modularity i.e., maximizing the edges within each community and minimizing the edges between communities, and we compare them with the locksteps detected by \system. We use the Python package \verb!igraph!\footnote{\url{http://igraph.org/}} to run these algorithms. 

\begin{table}[t]
\centering
\caption{Community detection and locksteps.}
\label{table:comm}
\scalebox{0.9}{
\begin{tabular}{l|l|l|}
\cline{2-3}
                                                         & FastGreedy \cite{fastgreedy} & Multilevel \cite{multilevel} \\ \hline
\multicolumn{1}{|l|}{Number of Communities}              & 6919       & 6439       \\ \hline
\multicolumn{1}{|l|}{Average \#nodes/community}          & 21         & 22         \\ \hline
\multicolumn{1}{|l|}{Median \#nodes/community}           & 2          & 2          \\ \hline
\multicolumn{1}{|l|}{Average \#locksteps/community}      & 2042       & 2387       \\ \hline
\multicolumn{1}{|l|}{Median \#locksteps/community}       & 7          & 31         \\ \hline
\multicolumn{1}{|l|}{Average Lockstep Coverage}          & 89.7       & 85.9       \\ \hline
\multicolumn{1}{|l|}{Median Lockstep Coverage}           & 91.67      & 87.5       \\ \hline
\multicolumn{1}{|l|}{Average \#Unique rep-pubs/community} & 9          & 11         \\ \hline
\end{tabular}}
\vspace{-10pt}
\end{table}

Table \ref{table:comm} shows the comparison of these algorithms with \system. 
Most of the communities are very small ($<3$ nodes). We observe that a large portion of the locksteps get mapped to the larger communities. The number of locksteps/community and the number of nodes/community reflect long tail distributions.
We define the lockstep coverage as the fraction of locksteps that reside within a single community. We predominantly observe locksteps having large ($>80\%$) coverage. Further, the number of unique rep-pubs per community is considerably large ($10$). This suggests that most of the communities are mixed up with locksteps coming from different publishers. This makes it difficult to logically assign each community to a particular group.
Community detection algorithms do not account for the timing of downloads, which makes it hard to pinpoint coordinated behavior between nodes.

\topic{Prior Lockstep Detection Algorithm}
We compare the locksteps detected by our algorithm to locksteps detected by the serial implementation of the CopyCatch \cite{copycatch} algorithm over one month (January 2013) of data. 
We reimplement CopyCatch, as the code is not available. 
There are qualitative differences between our algorithm and CopyCatch. 
Firstly, our algorithm is unsupervised. In contrast, CopyCatch requires seed domains corresponding to malicious domains and also times for all the domains at which some suspicious activity has occurred. 
Secondly, given a batch of data, we detect all the locksteps within that batch; CopyCatch can detect one single lockstep, which depends to the seed. 
Thirdly, CopyCatch solves an optimization problem to detect locksteps, which makes it highly sensitive to the choice of seed domains and the times provided. Furthermore, this serial implementation of CopyCatch is not scalable for large lockstep sizes; we consider only small locksteps for comparison.

To make a fair comparison, we generate 470 locksteps using our algorithm over the one month data. Of these only 139 locksteps have a size less than 10 which we consider for comparison. For each lockstep our algorithm detected, we provide CopyCatch the domains as seed nodes and the timestamp at which each domain was active in the lockstep as the seed times. Our algorithm generates 470 locksteps in $7.56$ s, taking an average of $0.016$ seconds per lockstep. In contrast, CopyCatch takes $600.9$ s to generate $139$ locksteps---an average of $4.32$ s per lockstep detection. 
These results suggest that \system shows promise for processing streaming data.

\vspace{-3pt}
\subsection{Robustness to evasion attempts}
\label{sec:adversarial-interference}
\vspace{-3pt}
An adversary could pursue three strategies for evading \system; we start by explaining these attacks in the context of $type_{dlr:dom}$ lockstep detection.
First, the adversary could frequently update or repack the downloaders it controls, so that no downloader is active in different time windows. 
This attack would prevent lockstep detection, and many malware families already employ aggressive repacking rates to evade detection~\cite{caballero2011measuring}. 
However, this strategy might impose a trade-off for organizations that conduct silent delivery campaigns, as they try to render their downloaders inconspicuous, e.g. by utilizing code signing and by avoiding behaviors that are not commonly seen in benign downloaders such as software updaters~\cite{Kwon15:DropperEffect}. 
The frequent updates and the lower prevalence of individual hashes that would result from higher repacking rates would make these downloaders look suspicious to an AV product. 
Instead of increasing the number of downloaders, in the second strategy the adversary could utilize a large number of domains, e.g. from DynDNS or a similar provider, so that each downloader accesses a single domain within a time window. 
This would be expensive for the adversary, as generating and registering new DNS domains is more costly than repacking downloaders and payloads. 
For example, to protect 500 droppers from lockstep detection, an adversary would need 5,000 DynDNS zones each month (\system considers second-level domains rather than FQDNs), at a current cost of \$4,000/month.\footnote{\url{http://dyn.com/managed-dns/}}
Additionally, this approach would make the domains more likely to be detected by DNS reputation systems, which use domain popularity as feature~\cite{bilge2014}.
In practice, \system detects MDLs that churn through more than 7 domains per day, as discussed in Section~\ref{sec:mdl-properties}.
The adversary could reduce the cost by instructing each downloader to randomly select a domain, from a pool of available domains, and to contact only that domain for $\Delta t$; then, the downloader would select another domain, and the reuse rate of domains in the pool would increase. 
To detect this, we could increase $\Delta t$, to cover the point when the downloader switches domains, and this would in turn force the adversary to further increase the time interval when each downloader accesses a single domain. 
Ultimately, the adversary cannot increase this time interval indefinitely, as domains that serve malware eventually get blacklisted. 
Additionally, we observe that the first two attack strategies involve increasing the downloader churn and reducing the domain churn, for evading the detection of $type_{dlr:dom}$ locksteps; to evade the detection of $type_{dom:dlr}$, these actions should be reversed. 
This suggests that it is difficult for an adversary to avoid both types of lockstep detection simultaneously. 
Finally, in the third strategy, the attacker could exploit the filtering step in our lockstep detection algorithm, for example by ensuring that MDLs appear deeper in our FP tree. 
In this case, \system is still able to capture a subset of these locksteps at lower FP tree levels.

\vspace{-3pt}
\section{Streaming performance}
\label{sec:evaluation}
\vspace{-3pt}

\begin{figure*}[t]
\vspace{-10pt}
\centering
\includegraphics[width=58mm]{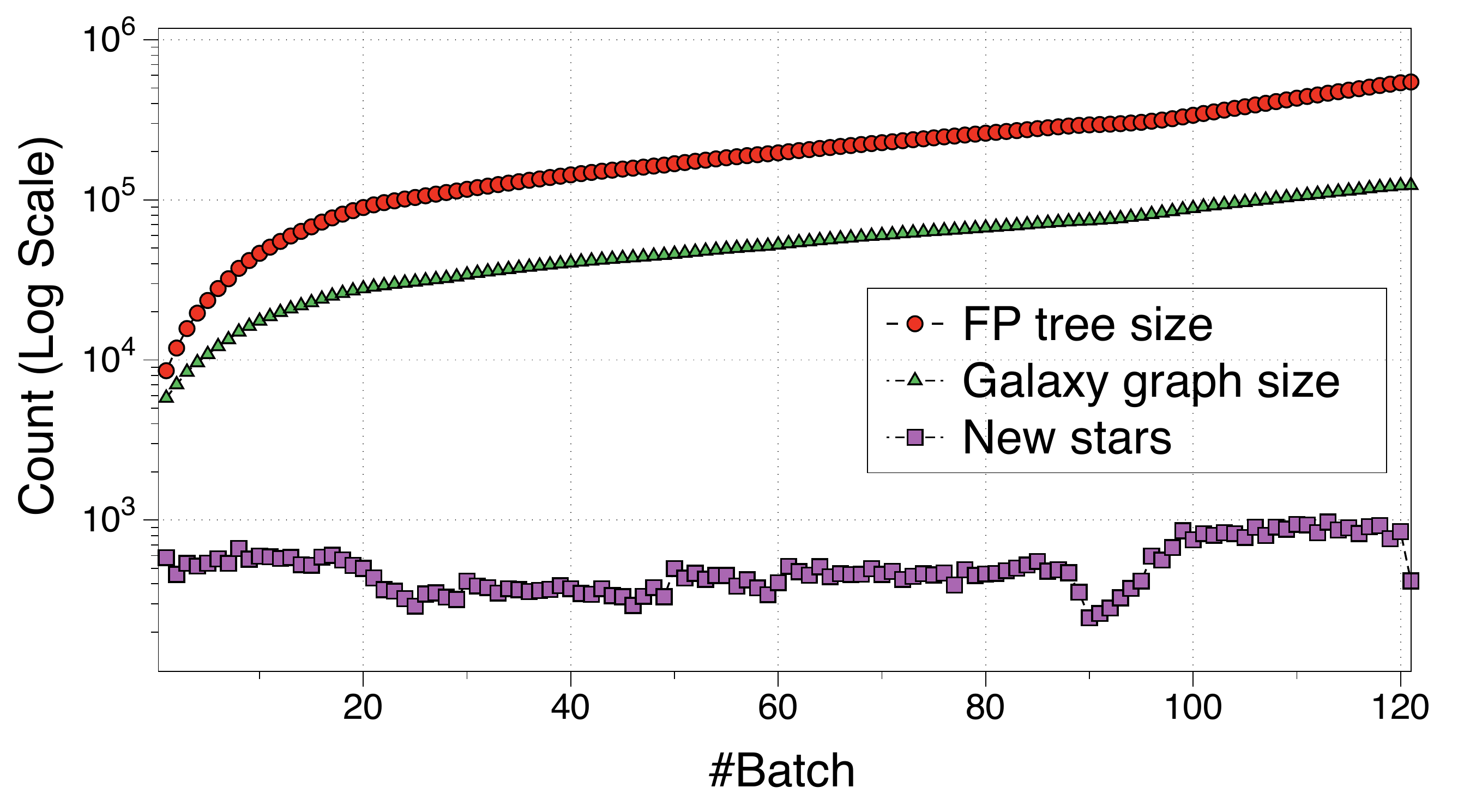}
\includegraphics[width=58mm]{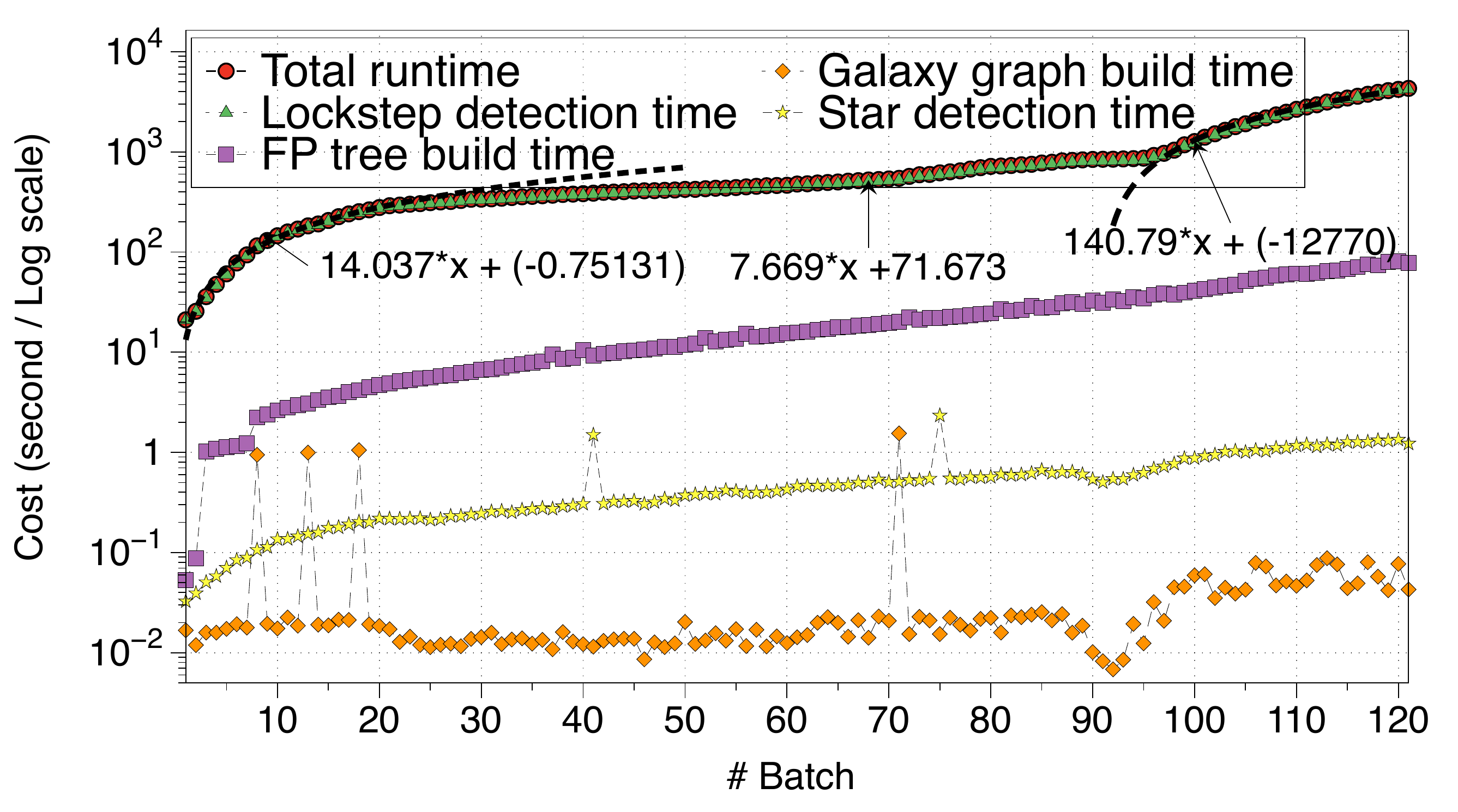}
\includegraphics[width=58mm]{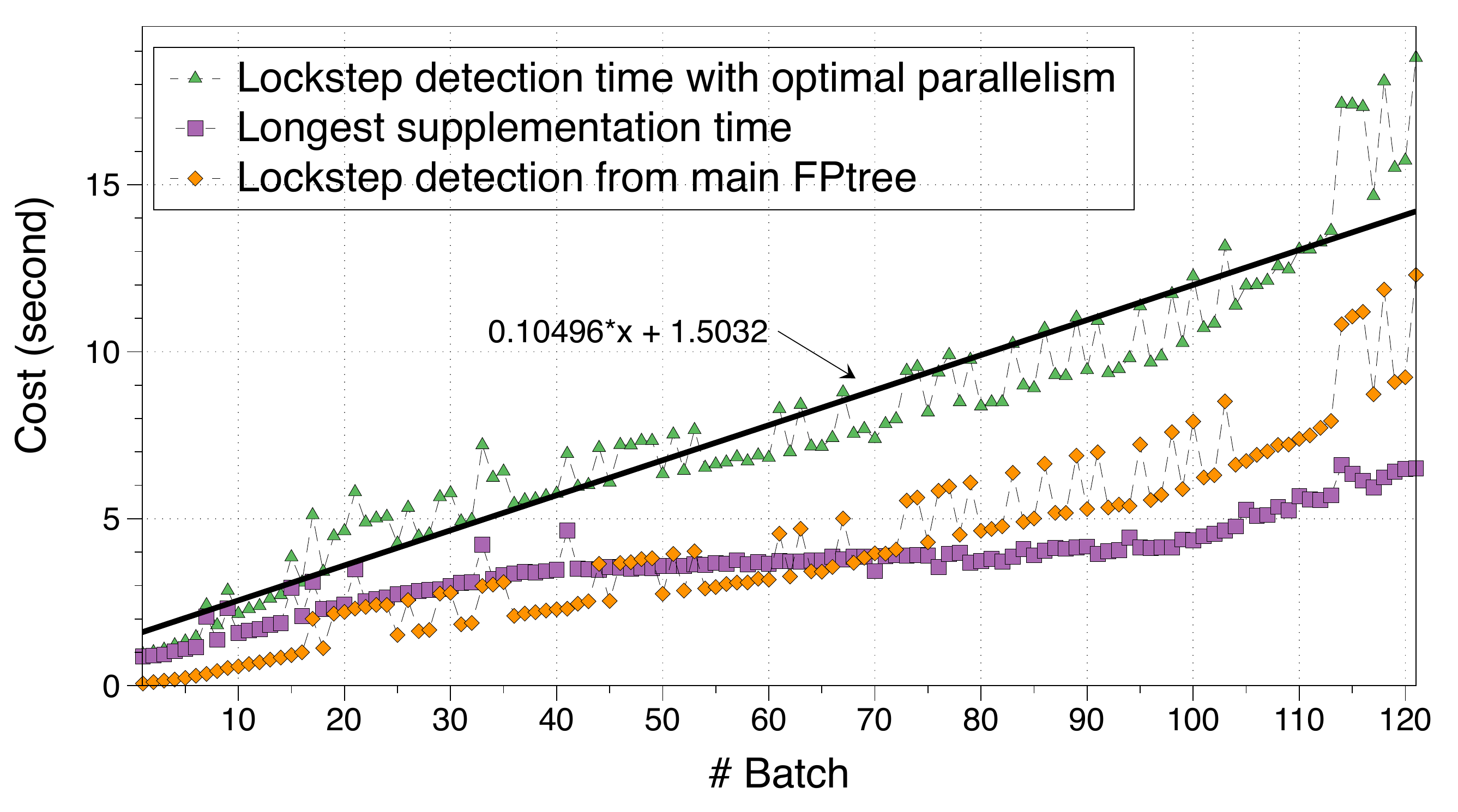}
\vspace{-5pt}
\caption{Streaming performance: (a) Data growth, (b) Running time of the streaming system, (c) Estimated lockstep detection runtime with optimal parallelism.} %
\vspace{-10pt}
\label{fig:streaming_result}
\end{figure*}

\topic{Experimental settings} 
We evaluate \system in streaming mode by feeding the download data in batches.  
In the lockstep detection phase, we filter out the FP tree level over \CutLevel, based on the observation that MDLs reside close to the root of the FP tree. 
And, we measure the latency of lockstep detection.
Each batch corresponds to a time window of $\Delta t = 3$ days.
As we employ one year of data, we have \NumDataPoints data points excluding the first batch in our experiment. %
For all \NumDataPoints data points, %
we measure the elapsed time for each of the four phases in our data analysis (illustrated in Figure~\ref{fig:fptree}).
We run our experiments on Amazon's Elastic Compute Cloud (Amazon EC2).\footnote{\url{https://aws.amazon.com/ec2/}} We use one M4.4xlarge instance, which has a 16-core 2.4 GHz Intel Xeon E5-2676 v3 (Haswell) with 64 GB of memory. 
For this evaluation, 
we focus on $type_{dlr:dom}$ graphs.

\topic{Streaming performance} 
Figure~\ref{fig:streaming_result}(a) illustrates the growth of the data structures that \system maintains. 
The plots has a logarithmic Y-axis, to compare both the number of new stars per batch and the cumulative number of nodes in 
the \graph and the FP tree.
On average, a batch contains 225,939 download events. 
Both the number of nodes in the galaxy graph and the FP tree grow linearly. 
At the end, the graph has 123,335 nodes and 637,814 edges.
As the data grows, the cost for detecting lockstep also grows incrementally. %

Figure~\ref{fig:streaming_result}(b) suggests that \system's runtime is dominated by the lockstep detection phase,  
which accounts for 97.2\% of the total runtime on average.
The total runtime shows three growth patterns: 
a steep increase for the first 20 batches,
a slower increase for most of the period, and 
another steep increase starting around batch 94--96.
Each of these growth patterns is linear and follows a regression line with the coefficients shown in the figure.
To further understand the latency of the lockstep detection step, recall that this phase consists of two parts: (1) lockstep detection on the main FP tree, (2) supplementation for finding partially missing locksteps (see Section~\ref{sec:lockstep-detection}). The near-biclique detection is done during lockstep detection, and it results in an overhead of at most 10 seconds.
As shown in Figure~\ref{fig:streaming_result}(c), the first part is fast, and requires at most 12 s. 
Most of the cost of lockstep detection comes from the supplementation effort, which 
induces the three phases of linear growth.
In particular, the number of nodes that have multiple versions in the FP tree increases significantly around batch 94--96, which triggers the third growth pattern in the total runtime. 

While \system searches for these nodes sequentially, we note that this could be done in parallel, as the supplementation sub-processes are independent of each other. 
To evaluate this potential optimization, we estimate the lockstep detection time with optimal parallelism. 
Assuming that enough computing resources are available for running all missing lockstep searches in parallel, the cost of this part of the computation will be determined by the longest running supplementation.
We obtain the total cost of lockstep detection with optimal parallelism by adding this to the runtime of lockstep detection on the main FP tree. 
As shown in Figure~\ref{fig:streaming_result}(c), this cost is at most 19 seconds, and shows a single pattern of slow linear growth. 
The supplementation phase is important for detecting malicious locksteps: 
at the last batch, this phase contributes to 95\% of the MDL detections and 91\% of the PDL detections. These locksteps include 48.7\% of the MDs and 80.6\% of the PDs.

Overall, these results suggest that the cost of \system's first two analysis steps is amortized over time, as we perform star detection only on the new batch of data and we maintain the \graph graph incrementally. 
The FP tree construction algorithm is not incremental and requires traversing the entire graph, but we optimize this step by pruning the FP tree at level \CutLevel, 
as we do not typically observe MDLs below this level. 
Similarly, the lockstep detection requires traversing the whole FP tree and constructing version lists for its nodes, but we could optimize this step by performing the supplementation in parallel. 
The resulting runtime of \system increases linearly with the size of the graph. 
Our results suggest that maintaining one year of download events imposes reasonable resource and performance requirements, even if we execute lockstep detection every day.

%% file: related_arxiv.tex
\vspace{-3pt}
\section{Related work}
\label{sec:related-work}
\vspace{-3pt}
\topic{Graph-based attack detection}
Zhao et al.\cite{zhao2009botgraph} introduces BotGraph that detects email accounts involved in spamming. They exploit the fact that botnet accounts share similar IP address and build a user-user graph. The aggressive sign-up behavior forces the botnet accounts to form a large cluster within the graph. 
Several works developed a reputation score system by adopting belief propagation, based on the intuition of locality.
Chau et al.\cite{chau2010polonium} exploit the tendency of hosts with poor cyber-hygiene having more malware. They construct a bipartite graph that represents the hosts and the files that present on those hosts.
Observing that several malware are distributed together, Tamersoy et al.\cite{tamersoy2014guilt} design a graph with files as nodes where edge is placed between the nodes that share a common host.
Oprea et al.\cite{oprea2015dsn} builds a host vs domain graph incrementally (day-by-day), and detects malicious domains within a same campaign.     
In \system, we maintain a graph based on the accessed by relationship between downloader and domain. The lockstep behavior detection returns clusters of downloaders and domains considering the temporal bounds.

\topic{Malware distribution} 
Cova et al.\cite{DBLP:conf/raid/CovaLTKD10} analyzed the rogue anti-virus campaigns by investigating the malicious domains involved in the distribution, introduced an attack attribution method employing feature-based clustering.
Vadrevu et al.\cite{Vadrevu13:MalwareDownloads} introduced AMICO, which is a system for detecting malware delivery in the live network traffic. They employed a supervised technique to classify malware download activities.
Invernizzi et al.\cite{Invernizzi14:Nazca} conducted the study on how the malware gets delivered through networks, proposed Nazca, a system that for detecting malicious download events from the web traffic.
Zhang et al.\cite{DBLP:conf/icdcs/ZhangSGLM15} employed unsupervised technique to identify the group of related severs that are likely to be involved in the same malware campaign. 
Contrary to these works, we conduct the study solely focusing on the client side of malware distribution networks, and employ unsupervised technique not based on features but on graph patterns. Another difference is in the way we attribute campaigns. While prior work generally relied on the properties of the malicious domains, we take advantage of the code signing behavior of the downloaders.

\topic{Spam campaigns}
Campaigns have been observed in other attack domains, for example in the context of spam. 
Several studies focused on email spam~\cite{DBLP:conf/ccs/KanichKLEVPS08,DBLP:conf/sp/LevchenkoPCEFGHKKLMWPVS11}
for example to measure the conversion rates and to analyze the resources involved in spam monetization.  
Spam campaigns have also been observed on social media sites~\cite{DBLP:conf/imc/GaoHWLCZ10,DBLP:conf/ccs/GrierTPZ10}. 
Prior work utilized machine learning techniques to characterize social media spam campaigns.
Some of the the prior techniques discussed use domain specific features that cannot be applied on the problem we are focusing on. However, the lockstep detection algorithm 
has broad applicability.

\topic{Lockstep detection}
CopyCatch \cite{copycatch} deals with identifying locksteps by analyzing the connectivity between users and pages through the likes relationship. We discuss the limitations of this algorithm and provide a comparison with \system in Section \ref{sec:prior-techniques-comparison}.
Most of the work in this space looks at detecting suspicious nodes \cite{catchsync} or suspicious edges \cite{autopart} through the lens of outlier detection. SynchroTrap \cite{SynchroTrap} proposes a malicious account detection system in the context of social networks to uncover malicious accounts and campaigns. They cluster users based on the Jaccard similarity of their actions. Our work is orthogonal to these techniques. Firstly, \system focusses on detecting malicious campaigns which correspond to near bipartite cores. Secondly, our system captures malicious campaigns over a large time interval; the notion of frequent patterns directly allows us to capture suspicious behavior. Finally, our algorithm is unsupervised. 

%% file: conclusions_arxiv.tex
\vspace{-3pt}
\section{Conclusions}
\label{sec:conclusions}
\vspace{-3pt}
We introduce \system, a system for systematically detecting silent delivery campaigns. 
\system detects lockstep behavior, which captures a set of downloaders that are controlled remotely and the domains that they access.
Using \system, we identify and analyze \NumCampaignsMillions million campaigns conducted in 2013.
We describe novel findings about malware distribution campaigns, such as an overlap between the malware and PUP delivery ecosystems and the tight business relationships among several PPI providers. 
We identify several properties of malware distribution locksteps, including their size, life cycle, and frequent domain changes, which 
allow us to implement several optimizations for detecting malware delivery campaigns in a streaming fashion. 
We also evaluate the performance of \system in streaming mode, and we show that 
it scales to large volumes of data.

%% file: pseudocode.tex
\newpage
\appendix[Pseudocode of \system]

\begin{algorithm}
  \caption{Star Detection.}
  \label{algorithm:star_detection}
  \begin{algorithmic}[1]
  	\Require Download events within $\Delta t$
    \Procedure{Detect New Stars}{}
      \BState \emph{Aggregate by $dom$}:
	  \For{Each event in Download events}
	  	\If{$dom$ is first seen}
	  		\State Initialize $dom.dlr\_list$
	  	\EndIf
	  	\If{$dlr \notin dom.dlr\_list$}
			\State Append $dlr$ to $dom.dlr\_list$
		\EndIf
	  \EndFor
	  \BState \emph{Assign star ID and get new stars}:
	  \State Initialize $new\_stars$
	  \For{Each $dom$}
	  	$star \gets (dom,dom.dlr\_list)$
	  	\If{$star \notin $ \texttt{stars}}
	  		\State Insert $(star\_id,star)$ to \texttt{stars}
	  		\State Append $(star\_id,star)$ to $new\_stars$
		\EndIf
	  \EndFor\\
	  \Return $new\_stars$
    \EndProcedure
  \end{algorithmic}
\end{algorithm}

\begin{algorithm}
  \caption{Galaxy Graph Maintenance.}
  \label{algorithm:galaxy_graph_maintenance}
  \begin{algorithmic}[1]
  	\Require $new\_stars$, G: \graph graph, G=(U,V,E) \Comment{$U,V \in nodes_{gg}$, E: edges}
    \Procedure{Update galaxy graph}{}
      \BState \emph{Check set relationship among stars}:
	  \For{$star$ in $new\_stars$}
	  	\State $star\_id,dom,dlr\_list \gets star$
	  	\Comment{Each $v \in V$ consists of $dom,version$}
	  	\State Initialize $is\_subset \gets 0$
	  	\For{Each $v$ with $dom=v.dom$}
	  		\If{$dlr\_list \in v.neighbors$}
	  			\State $is\_subset \gets 1$
	  			\State \textbf{break}
	  		\EndIf
	  		\If{$v.neighbors \in dlr\_list$}
	  			\State Flag $v$ as $to\_remove$
	  		\EndIf
	  	\EndFor
	  	\If{$is\_subset!=1$}
	  		\State Remove $v$s at $to\_remove$ from $\{V\}$
	  		\State Add $v\_new \gets (dom,star\_id)$ to $\{V\}$
	  		\For{$dlr \in dlr\_list$}
	  			\State Add edge $dlr,v\_new$
	  		\EndFor
	  	\EndIf
	  \EndFor\\
	  \Return G
    \EndProcedure
  \end{algorithmic}
\end{algorithm}

\begin{algorithm}
  \caption{FP Tree Construction.} 
  \label{algorithm:fptree_construction}
  \begin{algorithmic}[1]
  	\Require G: \graph graph, G=(U,V,E)
    \Procedure{Build FP Tree}{}
      \BState \emph{Sort adjacent list of V}:
      \State Initiate $node\_version(u,0)$
      \State Sort $v \in \{V\}$ by $v.degree (descending)$
	  \For{$v \in \{V\}$}
		\State Sort $u \in v.neighbors$ by $u.degree (descending)$
	  \EndFor
	  \BState \emph{Insert nodes into FP Tree}:
	  \For{$v \in \{V\}$}
      	\State $curr\_node \gets root$
		\For{$u \in v.neighbor$}
		  \State $node\_version[u]++$
		  \If {$u \notin curr\_node.children$}
		    \State Add $u$ as child of $curr\_node$
		    \State Initialize $curr\_node.visited$
		  \EndIf
		  \State Add $v$ to $curr\_node.visited$
		  \State $curr\_node \gets u$
		\EndFor
	  \EndFor\\
	  \Return FPtree, $node\_version$
    \EndProcedure
  \end{algorithmic}
\end{algorithm}

\begin{algorithm}
  \caption{Lockstep Detection.}
  \label{algorithm:lockstep_detection}
  \begin{algorithmic}[1]
  	\Require FPtree, $\alpha_{min}$, $node\_version$
    \Procedure{Extract Locksteps}{}
      \BState \emph{Check set relationship among stars}:
      \BState Initialize $visit\_list \gets ['root']$
      \BState Initialize $locksteps$
      \While{$visit\_list.length > 0$}
      	\State $curr\_node \gets visit\_list[0]$
      	\State $visit\_list \gets visit\_list + curr\_node.children$
      	\If{$alpha < 1.0$}
      		\State $uset,vset \gets GET\_NEAR-BICLIQUES$
      	\Else
      		\State $uset,vset \gets path\_to\_curr\_node, curr\_node.visited$
      	\EndIf
      	\If{$Filter\_check=False$}
      		\State Append $(uset,vset)$ to $locksteps$
      	\EndIf
      	\State 
      \EndWhile
      \For{$u$ with $node\_version > 1$} \Comment{Supplementation phase}
      	\State Prepare $stars$ having $u$
      	\State Apply all procedures using $stars$ 
      	\State $locksteps \gets locksteps+locksteps_{supplement}$ 
      \EndFor\\
      \Return $locksteps$
    \EndProcedure\\
    \hrulefill
    \Procedure{$Filter\_check$}{}
    	\If{$uset.length < 3 || vset.length < 3$}
    		\Return $True$ 
    	\EndIf
    	\For{$child \in curr\_node.children$}
    		\If{$curr\_node.visited = child.visited$}
    			\Return $True$
    		\EndIf
    	\EndFor\\
    	\Return $False$
    \EndProcedure
  \end{algorithmic}
\end{algorithm}

\begin{algorithm}
  \caption{Near-bicliques}
  \label{algorithm:near_bicliques}
  \begin{algorithmic}[1]
    \Procedure{Get Near-bicliques}{}
    	\State Initialize $candidate\_list(key,value)$
    	\State Initialize $missing_v \gets 0$
    	\State $parent\_node \gets curr\_node.parent$
    	\While{$parent\_node != root$}
    		\State $missing_v++$
    		\If {$parent\_node.visited.length > curr\_node.visited.length$}
    			\For{$v \in parent\_node.visited-curr\_node.visited$}
    				\State Append $(v,missing_v)$ to $candidate\_list$
    			\EndFor
    			\State \textbf{break}
    		\EndIf
    	\EndWhile
    	\For{$child \in curr\_node.children$}
    		\State $missing_u \gets (curr\_node.visited-child.visited).length$
    		\State Append $(child,missing\_u)$ to $candidate\_list$
    	\EndFor
    	\State Sort $candidate\_list$ by $value (ascending)$
    	\State Add $node$ from $candidate\_list$ while $\alpha > \alpha_{min}$
    	\Return Updated $uset,vset$
    \EndProcedure
  \end{algorithmic}
\end{algorithm}